\documentstyle[emulateapj]{article}

\newcommand{\arcm}{\ifmmode {^{\scriptstyle\prime}}
          \else $^{\scriptstyle\prime}$\fi}
\newcommand{\parcm}[1]{\ifmmode {^{\scriptstyle\prime}\!.#1}
          \else $^{\scriptstyle\prime}\!.#1$\fi}
\newcommand{\muK}{\,$\mu$\hbox{K}}
\newcommand{\entry}[4]{#1, {#2}, {#3}, {#4}.}
\newcommand{\bentry}[3]{#1, {#2} (#3).}
\newcommand{\dtrms}{82^{+12.1}_{-9.1}~\mu{\rm K}}
\newcommand{\dtband}{59^{+8.6}_{-6.5}~\mu{\rm K}}

\slugcomment{}
\lefthead{Leitch et al.}
\righthead{CMBR Anisotropy}
\begin{document}
\hyphenation{an-iso-tropy}
\hyphenation{lin-ear-ity}
\hyphenation{non-lin-ear-ity}
\hyphenation{cull-ing}
\hyphenation{wave-guide}
\hyphenation{brems-strah-lung}
\hyphenation{arc-min-ute}

\title{A Measurement of Anisotropy in the Cosmic Microwave Background
on 7\arcm-22\arcm~Scales}

\author{E.M. Leitch\altaffilmark{1}, A.C.S. Readhead, T.J. Pearson}
\affil{California Institute of Technology, 105-24\\Pasadena CA 91125}
\altaffiltext{1}{Present address: University of Chicago, 5640 S. Ellis Ave., Chicago IL 60637.}
\author{S.T. Myers\altaffilmark{2}}
\affil{University of Pennsylvania\\209 S. 33rd St./Philadelphia PA 19104.}
\altaffiltext{2}{Present address: National Radio Astronomy
Observatory, P.O. Box 0, Socorro, NM 87801}
\author{S. Gulkis, C.R. Lawrence}
\affil{Jet Propulsion Laboratory, California Institute of
Technology\\4800 Oak Grove Dr./Pasadena CA 91109}

\addtocounter{footnote}{-1}

\begin{abstract}
We report a measurement of anisotropy in the cosmic microwave
background radiation (CMBR) on $7\arcm-22\arcm$~scales.  Observations
of 36 fields near the North Celestial Pole (NCP) were made at 31.7 and
14.5 GHz, using the 5.5-meter and 40-meter telescopes at the Owens
Valley Radio Observatory (OVRO) from 1993 to 1996.  Multi-epoch VLA
observations at 8.5 and 15 GHz allow removal of discrete source
contamination.  After point-source subtraction, we detect significant
structure, which we identify with emission from a combination of a
steep-spectrum foreground and the CMBR.  The foreground component is
found to correlate with {\it IRAS} $100~\mu$m dust emission.  Lack of
H$\alpha$ emission near the NCP suggests that this foreground is
either high-temperature thermal bremsstrahlung ($T_e \gtrsim 10^6~{\rm
K}$), flat-spectrum synchrotron or an exotic component of dust
emission.  On the basis of low-frequency maps of the NCP, we can
restrict the spectral index of the foreground to $\beta \geq -2.2$.
Although the foreground signal dominates at 14.5 GHz, the extracted
CMBR component contributes 88\% of the variance at 31.7 GHz, yielding
an {\rm rms} fluctuation amplitude of $\dtrms$, including $4.3\%$
calibration uncertainty and $12\%$ sample variance ($68\%$
confidence).  In terms of the angular power spectrum, $C_l =
\langle|a_l^m|^2\rangle$, averaged over a range of multipoles $l =
361-756$, the detected broadband amplitude is $\delta T_{l_e} \equiv
[{l(l+1)C_l/2\pi}]^{1/2} = \dtband$.  This measurement, when combined
with small angular-scale upper limits obtained at the OVRO, indicates
that the CMBR angular power spectrum decreases between $l\sim600$ and
$l\sim 2000$ and is consistent with flat cosmological models.

\end{abstract}

\keywords{cosmic microwave background --- cosmology: observations}

\section{Introduction}

In standard cosmological scenarios, baryonic matter decouples from the
radiation field at $z_{dec}\simeq 1100$.  Thus, the horizon scale at
decoupling separates scales of importance for structure formation from
scales which probe only the primordial spectrum of perturbations. For 
$z \gg 1$, the Hubble radius subtends an angle ${\Delta{\theta}_H} \approx
{0^\circ\!.87}\,\Omega_0^{1/2}({z/1100})^{-1/2}$, so that for
$\Omega_0 = 1$, angles $\gtrsim 1^\circ$ correspond to physical scales
which were outside the horizon at decoupling (see, e.g., \cite{KT}).  On scales
$\lesssim 1^\circ$, anisotropies are directly linked to causal physical
processes in the early universe. Specifically, they record the
amplitude and phase of acoustic oscillations in the baryon-photon plasma and
as such provide a direct measure of the cosmological parameters which
govern the early universe.

In this paper, we present results from an experiment designed to
detect CMBR anisotropy on $7\arcm-22\arcm$~scales ($l\simeq 600$).
This ``RING5M'' experiment is the most recent in a series of
filled-aperture, ground-based anisotropy
experiments at the Owens Valley Radio Observatory (OVRO).  Previous
experiments at the OVRO include the NCP experiment (\cite{NCP}),
in which  the OVRO 40-meter telescope was used to place a 95\%
confidence upper limit of $\Delta T/T < 1.7\times 10^{-5}$ on power at
$2\arcm$~scales, and the RING40M experiment (\cite{Myersthesis}),
which resulted in a detection of anisotropy, attributed to foreground 
contamination, of $2.3\times 10^{-5} < \Delta T/T < 4.5\times 10^{-5}$
(95\% confidence) at the same resolution.

Since the {\sl COBE} detection of a CMBR quadrupole in 1992
(\cite{cobe}), a large number of experiments have reported detections
of anisotropy on $\gtrsim 1^\circ$ scales (see \cite{Hancock},
\cite{Bondrev} for recent reviews).  The RING5M is one of the few
experiments to probe the region of $l$-space between these
experiments and the high-$l$ range of the earlier OVRO work (recent
results from the CAT telescope (\cite{CAT2}) provide the only other
detection on comparable scales).

Section \S\ref{sec:experiment} of this paper provides an overview of 
the RING5M experiment, while \S\ref{sec:obs} and \S\ref{sec:rx}
review the data acquisition and relevant receiver characteristics in
greater detail.  Sections
\S\ref{sec:fluxscale} and \S\ref{sec:calibration} 
describe the calibration of both telescopes used in the RING5M
experiment; data selection and editing are described in
\S\ref{sec:data}.  Results are presented in
\S\ref{sec:data}-\S\ref{sec:rfi}, in which we also explore the
data for internal consistency and describe tests for possible
sources of systematic error.  The results of an 8.5 GHz VLA survey of
the RING5M fields and subsequent multi-frequency monitoring of point
sources for subtraction from the anisotropy data are presented in 
\S\ref{sec:ptsrc}, while contamination by Galactic foregrounds
is discussed in \S\ref{sec:foregrounds}.  Sections
\S\ref{sec:cmbr}-\S\ref{sec:band} describe the detected CMBR
anisotropy.  The paper concludes with a discussion of the
significance of our result for cosmological models.

\section{The Experiment}
\label{sec:experiment}

Observations were conducted at the OVRO from 1993 to
1996, using HEMT-amplified radiometers on the OVRO 5.5-meter and
40-meter telescopes, fully-steerable parabolic reflectors with
Cassegrain focus (5.5-meter) and primary focus (40-meter).  At 31.7 GHz, the
resolution of the 5.5-meter telescope is $7\parcm{37}$ (FWHM).  At
14.5 GHz, the OVRO 40-meter telescope provides a second frequency
channel, for spectral discrimination of foregrounds.  Since full
illumination of the 40-meter telescope at 14.5 GHz would produce a
$\sim2\arcm$~beam, the feeds were designed to illuminate an $\sim11$~m
patch on the dish surface, providing a good match to the the 5.5-meter
beam at 31.7 GHz (see Table~\ref{tab1}).  The 14.5 GHz receiver
is mounted in an off-axis configuration to minimize shadowing by the
prime focus cage as well as scattering from the prime focus support
legs, thereby reducing the effect of ground spillover.  This makes the
beamwidth a strong function of focus position and hence of zenith
angle; the beam can vary by as much as an arcminute over the full
zenith angle range.  As a result, all observations at 14.5 GHz were
restricted to lie within $\pm5^\circ$ of the observing zenith angle
for the Ring, $Z_{\rm Ring}\simeq 50^\circ$ (see below).

The 5.5-meter telescope is illuminated from the Cassegrain
focus, so that the largest sidelobes of the feed illumination pattern
see the sky instead of the ground, and the total contribution of the
ground to the system temperature at 31.7 GHz is 6 K.  Although the
14.5 GHz receiver is located at the prime focus of the 40-meter
telescope, the under-illumination of the 40-meter dish results in a
primary antenna pattern having its first sidelobes directed at the
sky, and the ground contributes $< 1~$K to the system temperature.

\begin{deluxetable}{lccc}
\footnotesize
\tablewidth{5.5in}
\tablenum{1}
\tablecaption{\label{tab1} \sc Parameters for the OVRO 5.5-m and 40-m Telescopes}
\tablehead{\colhead{\phm{0}} & \colhead{\phm{0}} & \colhead{5.5-meter} &
\colhead{40-meter}}
\startdata
Center Frequency (GHz)                         &$\nu_c$ &31.7 & 14.5\nl
Bandwidth (GHz)	                               &$\Delta\nu$	      &6		          & 3\nl
RMS Sensitivity ($\rm mK~s^{1/2}$)             &$\sigma_{\rm rms}$    &$1.4$& $2.3$\nl
Major Axis Beamwidth (FWHM)                    &$\theta_{\rm maj}$    &$\phm{\tablenotemark{a}}7\parcm{40}\pm0\parcm{26}\tablenotemark{a}$&$\phm{\tablenotemark{a}}7\parcm{80}\pm0\parcm{97}\tablenotemark{a}$\nl
Minor Axis Beamwidth (FWHM)	               &$\theta_{\rm min}$    &$\phm{\tablenotemark{a}}7\parcm{34}\pm0\parcm{25}\tablenotemark{a}$&$\phm{\tablenotemark{a}}7\parcm{17}\pm0\parcm{81}\tablenotemark{a}$\nl
Beamthrow	                               &$\Delta\phi$          &$22\parcm{16}$	    & $21\parcm{50}$\nl
Main Beam Solid Angle ($10^{-6}{\rm sr}$)      &$\Omega_m$	      &$5.21\pm0.03$ & $5.35\pm0.02$\nl
Beam Solid Angle ($10^{-6}{\rm sr}$)&$\Omega_a$&$7.92\pm0.28$         &$7.28\pm0.28$\nl
Beam Efficiency                                &$\eta_b$              &$0.658\pm 0.024$& $0.735\pm 0.029$\nl
Aperture Efficiency                            &$\eta_a$              &$0.476\pm0.017$ &\phm{\tablenotemark{b}}\nodata\tablenotemark{b} \nl
Sensitivity ($\rm mK~Jy^{-1})$                 &$\Gamma$              &$4.10\pm0.15$ & $21.30\pm0.82\phm{0}$ \nl
\enddata
\tablenotetext{a}{Parameters are for the average of the ANT and REF beams.}
\tablenotetext{b}{For the under-illuminated 40-meter
telescope, the physical aperture is not well determined (see \S\ref{sec:calibration}).}
\end{deluxetable}

On both telescopes, the receiver input is continuously switched at 500 Hz
between two feed horns separated by $\sim 22\arcm$ on the sky; the recorded
signal is the difference between successive millisecond integrations from
alternate feeds.  On the 5.5-meter telescope, the columns of air seen
by the two feeds overlap in area by $> 10\%$ to $400$~m, while on
the 40-meter, the columns overlap by $> 10\%$ to $1$~km, and the fast 
(``Dicke'') switching freezes out atmospheric fluctuations which occur in both
beams simultaneously, at the same time suppressing $1/f$ noise from
receiver components.  During each measurement, azimuthal nodding of
the telescopes between two symmetric positions, offset by the
22\arcm~beamthrow, provides an additional level of spatial
switching, removing constant offsets or linear temperature gradients
from the sky or ground (see \S\ref{sec:obs}).  This ``double
switching'' technique has been used successfully in both previous OVRO
anisotropy experiments.

\begin{deluxetable}{lcc}
\tablewidth{0pt}
\tablenum{2}
\tablecaption{\label{tab2}\sc Coordinates of the Ring5m Fields}
\tablehead{\colhead{Field} & \colhead{$\alpha({\rm J2000})$} &
\colhead{$\delta({\rm J2000})$}}
\startdata
{\sc OV5M0024} & {\sc 00:24:45.1} & {\sc  87:55:01} \nl
{\sc OV5M0104} & {\sc 01:05:25.9} & {\sc  87:54:55} \nl
{\sc OV5M0144} & {\sc 01:46:04.6} & {\sc  87:54:47} \nl
{\sc OV5M0224} & {\sc 02:26:40.0} & {\sc  87:54:36} \nl
{\sc OV5M0304} & {\sc 03:07:10.9} & {\sc  87:54:20} \nl
{\sc OV5M0344} & {\sc 03:47:36.5} & {\sc  87:54:03} \nl
{\sc OV5M0424} & {\sc 04:27:55.9} & {\sc  87:53:43} \nl
{\sc OV5M0504} & {\sc 05:08:08.6} & {\sc  87:53:20} \nl
{\sc OV5M0544} & {\sc 05:48:14.3} & {\sc  87:52:59} \nl
{\sc OV5M0624} & {\sc 06:28:12.9} & {\sc  87:52:37} \nl
{\sc OV5M0704} & {\sc 07:08:04.4} & {\sc  87:52:14} \nl
{\sc OV5M0744} & {\sc 07:47:49.2} & {\sc  87:51:53} \nl
{\sc OV5M0824} & {\sc 08:27:27.8} & {\sc  87:51:33} \nl
{\sc OV5M0904} & {\sc 09:07:00.9} & {\sc  87:51:16} \nl
{\sc OV5M0944} & {\sc 09:46:29.2} & {\sc  87:51:02} \nl
{\sc OV5M1024} & {\sc 10:25:53.8} & {\sc  87:50:51} \nl
{\sc OV5M1104} & {\sc 11:05:15.7} & {\sc  87:50:43} \nl
{\sc OV5M1144} & {\sc 11:44:35.9} & {\sc  87:50:41} \nl
{\sc OV5M1224} & {\sc 12:23:55.7} & {\sc  87:50:41} \nl
{\sc OV5M1304} & {\sc 13:03:16.2} & {\sc  87:50:45} \nl
{\sc OV5M1344} & {\sc 13:42:38.5} & {\sc  87:50:53} \nl
{\sc OV5M1424} & {\sc 14:22:03.8} & {\sc  87:51:06} \nl
{\sc OV5M1504} & {\sc 15:01:33.1} & {\sc  87:51:20} \nl
{\sc OV5M1544} & {\sc 15:41:07.2} & {\sc  87:51:37} \nl
{\sc OV5M1624} & {\sc 16:20:47.1} & {\sc  87:51:57} \nl
{\sc OV5M1704} & {\sc 17:00:33.3} & {\sc  87:52:18} \nl
{\sc OV5M1744} & {\sc 17:40:26.2} & {\sc  87:52:41} \nl
{\sc OV5M1824} & {\sc 18:20:26.2} & {\sc  87:53:03} \nl
{\sc OV5M1904} & {\sc 19:00:33.4} & {\sc  87:53:26} \nl
{\sc OV5M1944} & {\sc 19:40:47.6} & {\sc  87:53:47} \nl
{\sc OV5M2024} & {\sc 20:21:08.8} & {\sc  87:54:07} \nl
{\sc OV5M2104} & {\sc 21:01:35.1} & {\sc  87:54:24} \nl
{\sc OV5M2144} & {\sc 21:42:07.1} & {\sc  87:54:38} \nl
{\sc OV5M2224} & {\sc 22:22:43.2} & {\sc  87:54:49} \nl
{\sc OV5M2304} & {\sc 23:03:22.5} & {\sc  87:54:57} \nl
{\sc OV5M2344} & {\sc 23:44:03.1} & {\sc  87:54:53} \nl
\enddata
\end{deluxetable}

In the RING5M experiment, we observe 36 fields spaced by the
22\arcm~beamthrow in a ring around the North Celestial Pole
(NCP).  Field positions are given in Table~\ref{tab2}.  In
order to suppress variations in the observed differential ground
temperature introduced by telescope motion, fields
are observed only within $\pm 5^\circ$ ($\pm 20^m$) of upper culmination
(transit, at $Z_{\rm Ring}\simeq 50^\circ$).  Near transit, the
separation of neighboring RING5M fields is approximately azimuthal, so
that if we denote the temperature in each field by $T_i$,
the quantity which results from the double switching is given by
\begin{equation}
\Delta T_i = T_i - {1\over 2}(T_{i-1} + T_{i+1}).
\label{eq:tring}
\end{equation}
The effective beam pattern produced by the switching is shown in 
Figure~\ref{fig1}.

\bigskip
\centerline{\vbox{\epsfxsize=8.0cm\center{\epsfbox{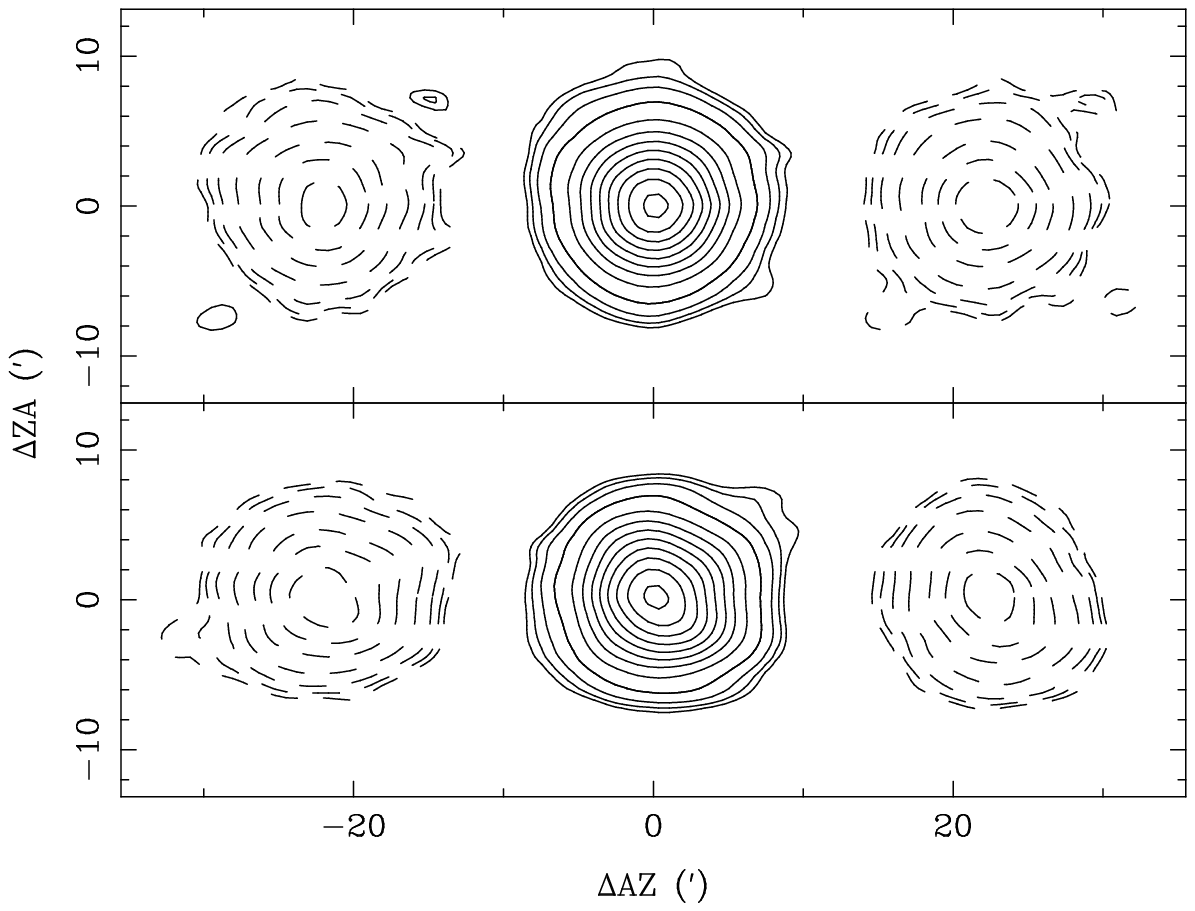}}
\figcaption{\footnotesize \label{fig1} Measured double-switched beam pattern at 31.7 GHz (top
panel) and 14.5 GHz (bottom), obtained by scanning across point
sources.  Contours are $-0.5$ to $-0.1$ in steps of $0.1$, $-0.05$,
$-0.03$, $-0.02$, $0.02$, $0.03$, $0.05$ and $0.1$ to $1.0$ in steps
of $0.1$.  For observations of the Ring, the symmetric negative beams 
correspond to the adjacent RING5M fields.}}}
\bigskip 

The interlocked geometry of the RING5M fields not only serves to eliminate
systematic differences between observations of different fields, 
but also provides a boundary condition
$\sum_i\Delta T_i = 0$ which can be checked for residual systematics.
Moreover, point sources in a RING5M field can be
identified by the characteristic minus-plus-minus signature they
produce in successive fields.

\section{Observations}
\label{sec:obs}

In all of the observations presented here, the telescope alternates
the beams (referred to as the ANT and REF beams) on source
by slewing in azimuth by an amount $\pm\Delta\phi$ equal to the
separation of the feed horns.  The telescope integrates for equal
times $\tau_s$ in each of four successive configurations, referred to
as the A, B, C and D fluxes, shown schematically in Figure~\ref{fig2}.
In combination with the fast differencing between the feeds ($\rm ANT
- REF$), this procedure, known as a ``FLUX'' procedure, forms the
basic double switching used to eliminate power gradients from the atmosphere or
ground  (see also \cite{NCP}, \cite{myers}).  The $\pm\Delta\phi$ positions are
referred to as {\it reference fields}.

To reduce systematic effects associated with varying slew times and
settling of the telescope structure (particularly on the
40-meter telescope), an adjustable idle time $\tau_i$ is inserted between the A
and B integrations and between the C and D integrations.  Because the
telescope does not move between the B and C integrations, no time delay is
inserted there.  The total duration of a FLUX procedure is thus $\tau
= 4\tau_s + 2\tau_i$.
For all of the RING5M observations presented here, $\tau_s = 20^s$ and
$\tau_i = 10^s$, so that a FLUX procedure typically requires $100^s$.
For each set of measurements, the quantity
\begin{equation}
{\rm FLUX} = {1\over{2}}(\Delta T_B+\Delta T_C-\Delta T_A-\Delta T_D)
\label{eq:fluxad}
\end{equation}
is formed, with associated standard deviation (SD) estimated by
summing the variances of the individual (A-D) integrations.

\bigskip
\centerline{\vbox{\epsfxsize=6.5cm\center{\hspace{0.25cm}\epsfbox{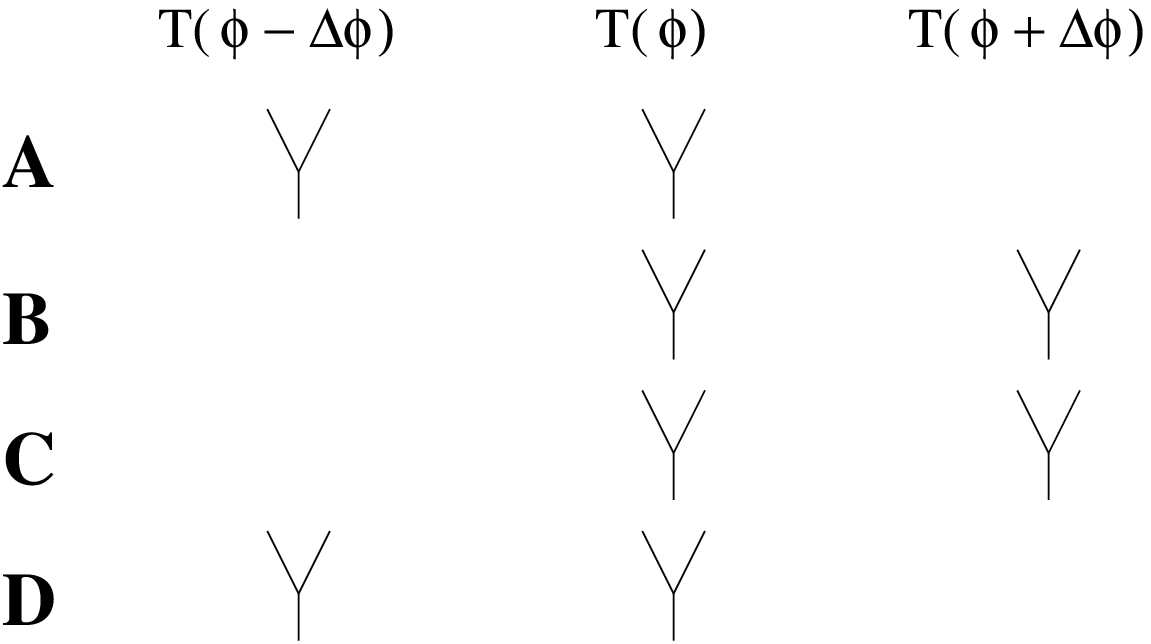}}
\figcaption{\footnotesize \label{fig2} Double switching pattern in a FLUX
procedure, shown here for an arbitrary temperature field on the sky.
The sense of the beams is ANT left, REF right.}}}
\bigskip 

From Figure~\ref{fig2}, it can be seen that the Dicke switching produces
\begin{eqnarray}
\nonumber
\Delta T_{A,D} &=& T(\phi-\Delta\phi)-T(\phi)\\
\Delta T_{B,C} &=& T(\phi)-T(\phi+\Delta\phi),
\label{eq:ad}
\end{eqnarray}
so that the double-switched FLUX is given by
\begin{eqnarray}
\label{eq:swdiff}
\nonumber
{\rm FLUX} &=& \left\{T(\phi) - T(\phi+\Delta\phi)\right\} \\\nonumber
& & \,\,\,\,-\left\{T(\phi-\Delta\phi)-T(\phi)\right\}\\
&\simeq& -{\Delta\phi}^2\left.{\partial^2 T\over{\partial\phi^2}}\right|_{\phi}.
\label{eq:flux}
\end{eqnarray}
(In Eq.~\ref{eq:tring}, $T_{i+1}$ and $T_{i-1}$ are the temperatures at
the $\pm\Delta\phi$ positions, respectively.)  If a source under
observation is smaller than the beamthrow of the telescope, i.e., 
$T_{\rm src}(\phi+\Delta\phi) = T_{\rm src}(\phi-\Delta\phi) = 0$, from
Eq.~\ref{eq:swdiff} we see that the contribution to the FLUX is just
$2T_{\rm src}(\phi)$, or twice the power increment that would be
measured with a single-difference observation.

On both telescopes, gain variations in the
amplifiers are removed by referencing to noise diodes.  Signals from
the diodes are injected just behind the feed horns (see
Figure~\ref{fig3}) and are
subject to the same receiver gain variations as the astronomical
signal.  The diode measurement, known as a ``CAL'' procedure, is
identical to the FLUX procedure just described, with the exception
that the telescope does not move between the A and B or the C and D 
integrations.  The diode remains off during the A and D integrations
and is turned on during the B and C integrations, so that from Eq. 
\ref{eq:fluxad}, the recorded CAL is just $T_{\rm diode}$.

\begin{deluxetable}{lcc}
\footnotesize
\tablewidth{4in}
\tablenum{3}
\tablecaption{\label{tab3}\sc Absolute Flux Density and
Temperature Scale}
\tablehead{\colhead{Source} & \colhead{31.7 GHz\tablenotemark{a}} & \colhead{14.5 GHz\tablenotemark{b}}}
\startdata
Jupiter	 & ${\bf{\phm{\rm K~}152\pm 5~{\rm K}\phm{12}}}$ & ${\phm{\rm
K~}175\pm 9~{\rm K}\phm{17}}$\nl
DR~21	 & $\phm{\tablenotemark{c}~{\rm Jy}}20.60\pm 0.68~{\rm
Jy}\tablenotemark{c}\phm{0}$ & ${\bf\phm{\tablenotemark{c}~{\rm
Jy}}22.87\pm1.07~{\rm Jy}\tablenotemark{c}\phm{1}}$ \nl
Cas A	 & $\phm{\tablenotemark{c}\phm{~{\rm Jy}}}164.18\pm 5.45~{\rm Jy}\tablenotemark{c}\phm{16}$ & $\phm{\tablenotemark{c}\phm{~{\rm Jy2}}}313.04\pm14.8~{\rm Jy}\tablenotemark{c}\phm{100}$ \nl
Crab	 & $\phm{\tablenotemark{c}\phm{~{\rm Jy}}}307.15\pm 10.14~{\rm Jy}\tablenotemark{c}\phm{3}$ &	$\phm{\tablenotemark{c}\phm{~{\rm Jy2}}}426.9\pm20.17~{\rm Jy}\tablenotemark{c}\phm{1}$ \nl  
3C~84	 & \nodata 			& $\phm{~{\rm Jy}}25.84\pm1.26~{\rm Jy}\phm{1}$ \nl
3C123	 & \nodata 			& $\phm{~{\rm Jy}}5.85\pm0.28~{\rm Jy}$ \nl
3C~218	 & \nodata 			& $\phm{~{\rm Jy}}4.69\pm 0.23~{\rm Jy}$ \nl        
3C~286	 & $\phm{~{\rm Jy}}2.02\pm 0.07~{\rm Jy}$ 	& $\phm{~{\rm Jy}}3.80\pm0.22~{\rm Jy}$ \nl         
3C~353   & \nodata			& $\phm{~{\rm Jy}}8.44\pm0.40~{\rm Jy}$ \nl         
NGC~7027 & \nodata			& $\phm{~{\rm Jy}}6.20\pm0.30~{\rm Jy}$
\enddata
\tablenotetext{a}{\parbox[t]{3.5in}{Assuming a brightness temperature
for Jupiter of $152\pm 5~$K at 31.7 GHz and $\Omega_{\rm Jup} = 6.656\times
10^{-7}~{\rm sr}$ at 1~au.}}
\tablenotetext{b}{Assuming a ratio of $S_{\rm DR~21}(14.5~{\rm GHz})/S_{\rm
DR~21}(31.7~{\rm GHz})$ given by Eq.~\ref{eq:dr21}.}
\tablenotetext{c}{\parbox[t]{3.5in}{Flux density as seen by the telescope beam; not an accurate measure of absolute flux density.}}
\end{deluxetable}

\section{Receiver Characteristics}
\label{sec:rx}

On the 5.5-meter telescope,
load switches identical to the Dicke switch provide an observing mode
where the background power is input from two internal loads of temperature
$T_L\simeq 20~$K, providing a stable background against which to
measure the noise diode powers even during periods of bad weather.  On
the 40-meter telescope, no internal loads are available, and all diode
measurements are performed against the sky.

Both receivers exhibit a small degree of non-linearity, i.e., the measured
power increment against a source in the presence of the typical
background power level underestimates the power increment at the
front of the feeds by $5-10\%$ (see \cite{Leitchthesis} for details).
As a result, care must be exercised when comparing noise diode
measurements against the internal
loads to observations against the sky, as the background
powers are in general different, leading to variations in the
FLUX/CAL ratio as large as 6\%.  On both receivers,
this effect has been measured and can be removed to high
accuracy; at both frequencies, the non-linearity correction
contributes $\lesssim 1\%$ to the final calibration error.

\bigskip
\centerline{\vbox{\epsfxsize=5.5cm\center{\hspace{0.25cm}\epsfbox{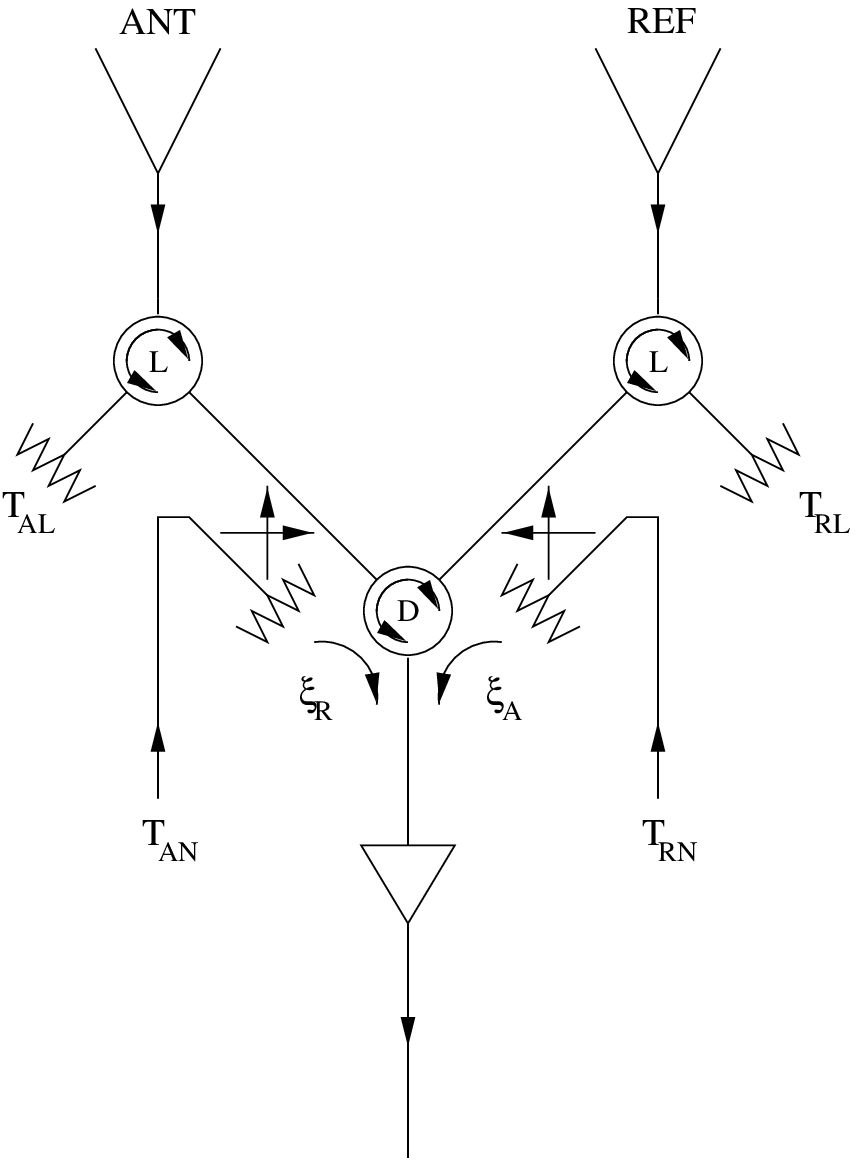}}
\figcaption{\footnotesize \label{fig3} A schematic of the 31.7 GHz receiver front end.  
$T_{\rm AL}$ and $T_{\rm RL}$ represent the contribution to the system
temperature of the ANT and REF internal loads, respectively.  Similarly, $T_{\rm
AN}$ is the temperature of the noise diode on the ANT side.  The
isolation of the Dicke switch (denoted D) when pointed at the ANT side is
indicated by $\xi_A$ and when pointed to the REF side, by $\xi_R$.}}}
\bigskip

The fast switching is accomplished by a Dicke switch --- a three-port,
wide-band circulator whose direction is determined by the
polarity of the magnetic field within its ferrite core.  With the
switch directed at one input port, a small amount of signal from the
other input port is transmitted to the output port, typically at the level of
$1-2\%$, known as the {\it isolation}, denoted $\xi$.  The directional
isolations of the switch, i.e., the isolations with the switch in ANT
or REF position, denoted $\xi_A$ and $\xi_R$ respectively, depend
sensitively on the impedance match at the three ports and in general
are not equal.

From Figure~\ref{fig2}, it can be seen that if $\xi_A\neq\xi_R$,
even a linear temperature gradient can partially survive the
double switching, contributing an additional term
\begin{equation}
\Delta {\rm FLUX} \simeq \Delta\phi{\partial 
T\over{\partial\phi}}(\xi_{\rm R} - \xi_{\rm A}),
\label{eq:grad}
\end{equation}
to Eq.~\ref{eq:flux}, which, given typical values of
$\xi_A$ and $\xi_R$, can be as large as
$100~\mu$K for sky gradients as small as $1~$mK/arcminute.  We believe this
effect to be responsible for year-to-year fluctuations in the mean
observed in the RING5M data (see \S\ref{sec:meanlevels}), which we
expect a priori to be zero (see \S\ref{sec:experiment}).

\section{Flux Density Scale}
\label{sec:fluxscale}

At 31.7 GHz, the flux density scale of the 5.5-meter telescope is
based on a 31.4 GHz measurement of the brightness temperature of Jupiter,
\begin{equation}
T_J = 152\pm5~{\rm K}
\label{eq:jup}
\end{equation}
 (\cite{Dent}).  During 1996, daily calibration of the
internal noise diodes at both frequencies was achieved by comparison
with a set of secondary standards whose flux densities
were measured relative to Jupiter (see Table~\ref{tab3}),
using an ephemeris distance for Jupiter and assuming $\Omega_{\rm Jup} =
6.656\times10^{-7}~{\rm sr}$ at 1 au.  Ratios to Jupiter were determined
at three epochs during which calibrator sources were observed for
several days, as well as from daily observations of the sources (see below).
Although the supernova remnants Cas~A and the Crab nebula are partially
resolved on both the 5.5-meter telescope and the under-illuminated
40-meter telescope ($\theta_{\rm snr}\sim4^\prime$), pointing on these sources
has proven reproducible to high precision, making them suitable as
relative calibrators.

Daily calibrator observations were interleaved with
observations of the RING5M fields, so that every twelfth field (every eight
hours) was replaced by a 40-minute scan on a calibrator source.  To
avoid selective depletion of data from any three fields, the set of secondary
calibrators was chosen so that at least one would be visible at any
time, and calibrator scans were precessed daily by one field, resulting in a
uniform reduction in sensitivity of only 4\% over the entire Ring.  
At 14.5 GHz, observations of calibrator sources were
restricted to lie within $\pm5^\circ$ of the zenith angle of the Ring,
and a correspondingly larger set of secondary calibrators was used to
satisfy this condition. 

\subsection{DR~21 and the 14.5 GHz Flux Density Scale}
\label{sec:dr21}

The 31.4 GHz brightness temperature of Jupiter in Eq.~\ref{eq:jup} is
based on a fit to the spectrum of the HII region DR~21,
\begin{equation}
S_\nu = 26.78 - 5.63 \log\nu_{\rm GHz}~{\rm Jy},
\label{eq:dr21}
\end{equation}
with an associated error of $\pm 3\%$ over the range $7-40$~GHz (\cite{Dent}).  
Because of its location in a complicated region of the Galactic plane,
DR~21 itself is not used as an absolute calibrator in this experiment.
Since the reference fields in a FLUX procedure are displaced
azimuthally by $22\arcm$ (see Figure~\ref{fig1}), as DR~21 is tracked
on the sky, emission in the ring of radius $22\arcm$ around the
source rotates through the reference beams, making the measured flux
density a function of the source parallactic angle $\psi_p$, defined
as the anglebetween the great circle passing through the source and
the zenith and the great circle passing through the source and the
celestial poles (throughout this paper, we have folded the parallactic
angle into the range ($-90^\circ, 90^\circ$)) (see Figure~\ref{fig4}).
Independent scans on DR21, however, agree over the
full parallactic angle range to within the scatter of the data.  The
resultant variation in the apparent flux density of DR~21 can
therefore be removed, making it suitable as a relative
calibrator.  

\bigskip
\centerline{\vbox{\epsfxsize=8.5cm\center{\epsfbox{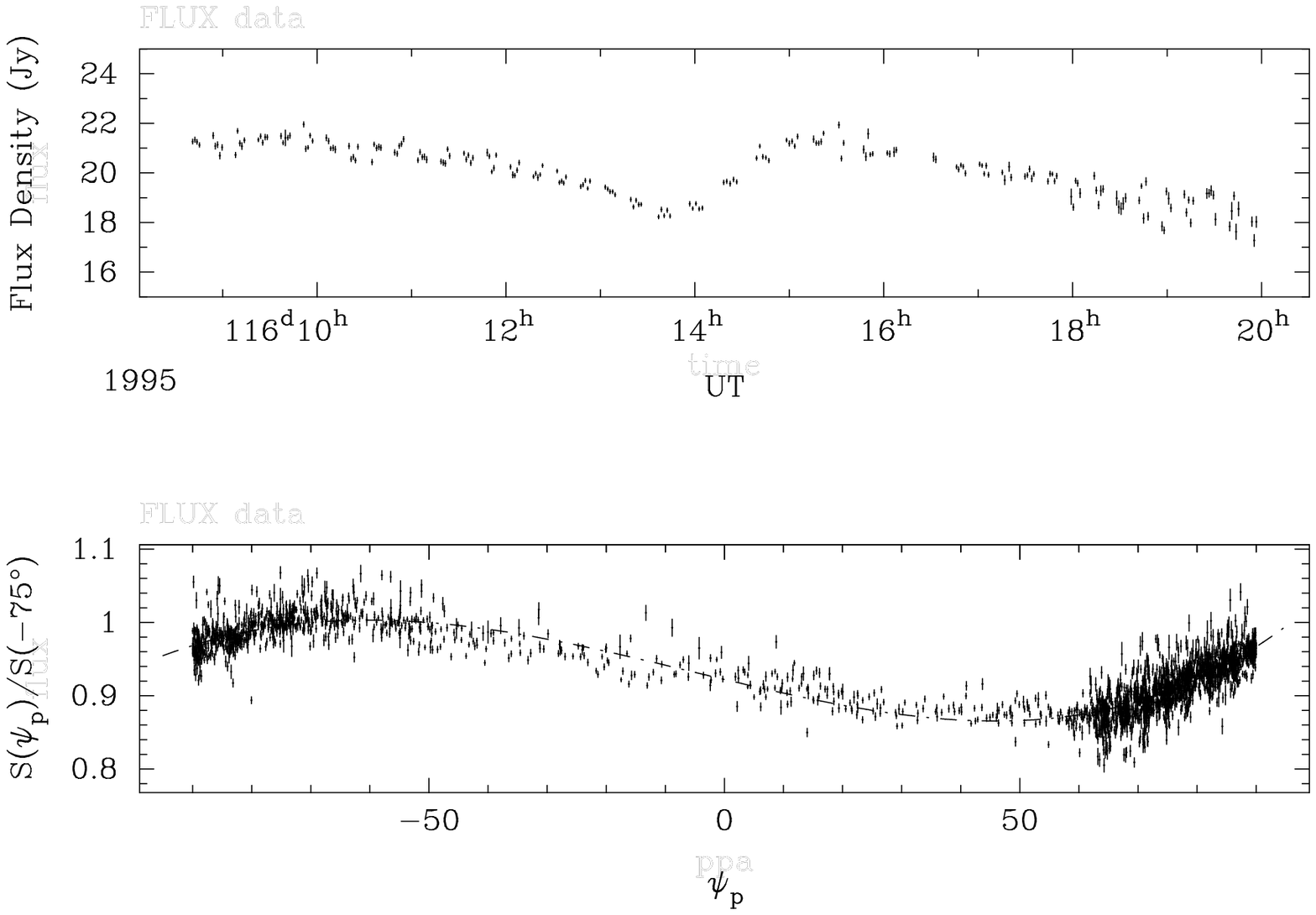}}
\figcaption{\footnotesize\label{fig4} (Top panel) A typical
scan on DR~21.  The apparent flux density is dominated by
contamination in the reference beams and is a strong
function of parallactic angle.  (Bottom) Dependence of the switched
flux on parallactic angle, shown here in a compilation of 34
independent tracks on DR~21 from 1995.  Each scan was separately calibrated and
normalized to the flux over the $\psi_p$ interval
($-80^\circ,-70^\circ$).  Shown also is the best fit model for the
$\psi_p$ dependence.}}}
\bigskip

It is found that the fit to $S_{\rm DR~21}(\psi_p)$ derived at 31.7
GHz also removes the parallactic angle dependence of the flux density
at 14.5 GHz, indicating that the contaminating flux in the
reference fields has the same spectrum as DR~21 between 31.7 and 14.5
GHz.  This is not surprising, since all of the emission in the region
surrounding DR~21 is thought to belong to the same HII complex.  Since
the frequency dependence of the contaminating flux is separable from
its parallactic angle dependence, the observed flux {\it
ratio} at the two frequencies should follow from Eq.~\ref{eq:dr21}, and
this fact can be used to establish a 14.5 GHz flux density scale
relative to the same measurement of Jupiter at 31.4 GHz.  The flux
density scale so derived is in excellent agreement with
the 14.5 GHz measurement of Jupiter's brightness temperature relative
to DR~21 obtained by \cite{Gary}, using the Goldstone 64-m telescope.
Therefore this scale, with the 5\% errors reported by Gary, is the one
used to calibrate the 14.5 GHz data.

\subsection{Cas A and Re-calibration of the Saskatoon Experiment}

The Saskatoon group has recently reported a measurement of intrinsic
anisotropy relative to a fit to the spectrum of Cas A (\cite{sask}).
The power they detect on degree scales has
led to speculation that their assumed flux density for Cas A may be too high.
Since Cas A is used as a calibrator source in the RING5M experiment,
the OVRO observations of Cas A not only provide a corroborative check
on the Saskatoon calibration, but can in principle refer the Saskatoon
measurements to a flux density scale based on Jupiter, thereby
reducing their calibration uncertainties to a few percent.  
Since Cas A is comparable in size to the 5.5-meter telescope beam, however, our
measurement cannot be compared directly with the Saskatoon value, but
must be multiplied by the factor  
\begin{equation}
f = {{\int_{\Omega_{\rm cas}}{P_{\rm
Cas}(\theta,\phi)d\Omega}}\over{\int_{\Omega_{\rm cas}}
{P_{\rm cas}(\theta,\phi)P_{\rm ovro}(\theta,\phi)d\Omega}}},
\label{eq:cascorr}
\end{equation}
where $P_{\rm cas}(\theta,\phi)$ is the source brightness distribution, and
$P_{\rm ovro}(\theta,\phi)$ is the normalized antenna power pattern,
given by the central lobe in Figure~\ref{fig1}.

The correction factor in Eq.~\ref{eq:cascorr} was determined from a 32
GHz map of Cas A, made with the 100-m Effelsberg telescope
(Morsi 1997, private communication), as a template for 
$P_{\rm cas}(\theta,\phi)$.  The effect of uncertainties in the
5.5-meter telescope pointing, as well as in the determination of the
telescope beam, was modeled via Monte Carlo simulation.
Gaussian-distributed pointing positions were generated for $10^5$
events, with $\sigma_{A} = \sigma_{Z} = 0\farcm{3}$ (typical for both 
telescopes) centered on the nominal pointing position; beamwidths were
drawn from Gaussian distributions centered on the best-fit
$\theta_{\rm min}$ and $\theta_{\rm maj}$.  The $68\%$ confidence
interval of the resulting distribution for $f$ is given by
\begin{equation}
f = 1.18^{+0.02}_{-0.01}.
\label{eq:correct}
\end{equation}
With the flux density as seen by the OVRO 5.5-meter telescope beam
given by $S^\prime_{\rm ovro} = 164.18\pm 5.45$~Jy (see Table~\ref{tab3}),
and accounting for the secular decrease in the flux density of Cas~A since 1994
(\cite{Baars}), application of Eq.~\ref{eq:correct} yields
\begin{equation}
S_{\rm ovro} = 195.59^{+7.29}_{-6.70}~{\rm Jy},
\end{equation}
or
\begin{equation}
S_{\rm ovro} = (1.05\pm 0.04)S_{\rm sask}
\end{equation}
(\cite{Netterthesis}), consistent with the Saskatoon calibration.

\section{Calibration}
\label{sec:calibration}

The power received from a source of specific intensity $I$ uniformly
filling the main beam $\Omega_m$ of the telescope power pattern is given by
\begin{equation}
P = {1\over{2}}IA_p\eta_a\Omega_m\Delta\nu,
\end{equation}
where $\eta_a$ is the ratio of the
effective aperture of a dish to its physical aperture $A_p$, known as
the {\it aperture efficiency}.  The power in
instrumental units is converted to physical units by comparison with the power
emitted by a calibrator source of known intensity $I_{\rm cal}$, typically a
noise diode internal to the receiver.  Since the 
radiation from an internal noise diode fills the beam
solid angle $\Omega_a$ of the telescope, the ratio $P/P_{\rm cal}$ is
given by 
\begin{equation}
{P\over{P_{\rm cal}}} = {{I\Omega_m}\over{I_{\rm cal}\Omega_a}}\equiv 
{{I\Omega_m}\over{S_{\rm cal}}},
\end{equation}
so that the intensity for a source filling the main beam is given by
\begin{equation}
I = \left(P\over{P_{\rm cal}}\right){S_{\rm cal}\over{\Omega_m}}.
\end{equation}
Atmospheric attenuation reduces the observed intensity of a source
by a factor $\kappa({Z}) = \exp(-\tau A(Z))$, where $\tau$ is the
atmospheric opacity at zenith and $A({Z})\simeq\sec{({Z})}$ is the
airmass.  Thus, to recover the intensity of a source above the
atmosphere, we must compute
\begin{equation}
I_0 = \left({P\over{P_{cal}}}\right)\left({S_{cal}\over{\Omega_m}}\right)
{1\over{\kappa({Z})}}.
\label{eq:cal1}
\end{equation}

Throughout this paper, we will alternately use {\it intensity $I$},
{\it brightness temperature $T_B$} (the equivalent Rayleigh-Jeans
(R-J) temperature of a source filling the main beam $\Omega_m$) and {\it antenna
temperature $T_A$} (the equivalent R-J temperature of a
source filling the beam solid angle $\Omega_a$).  These are related
simply by
\begin{equation}
I\Omega_m = {2kT_B\over{\lambda^2}}\Omega_m = {2kT_A\over{\lambda^2}}\Omega_a,
\label{eq:convert}
\end{equation}
whence
\begin{equation}
T_B = {T_A\over{\eta_b}},
\end{equation}
where $\eta_b \equiv \Omega_m/\Omega_a$.

The mean 31.7 GHz zenith atmospheric opacity at the OVRO during 1996
is determined by fits to the daily calibrator source observations, yielding
$\overline{\tau}_{31.7} = 0.045\pm0.002$, consistent with the mean
annual zenith opacity estimated from a water vapor radiometry (WVR) system at
the OVRO
during 1994-1996.  The error in $I_0$ introduced by adopting a
constant mean opacity during 1994-1996 is $< 0.02\%$ (\cite{Leitchthesis}).
Extrapolation to 14.5 GHz of WVR opacities measured at 31.4 and 20
GHz yields $\overline{\tau}_{14.5} = 0.023$.

Maps of the main beam, shown in Figure~\ref{fig1},
were obtained in 1995 from raster-scans across Jupiter and 3C~84 (for
the $7\parcm{4}$ beam, Jupiter is approximately a point source).  At 14.5
GHz, scans were restricted to lie within $\pm5^\circ$ of ${Z}_{\rm
Ring}$ so that the resulting beam map is the one appropriate
for calibration of the RING5M data (see \S\ref{sec:experiment}).  At
both frequencies, the main beam solid angle $\Omega_m$ is determined
to an accuracy of $\leq 1\%$.

As discussed in \S\ref{sec:fluxscale}, three independent
estimates of the noise diode flux density $S_{\rm cal}$ were obtained
each day during 1996,
from which we deduce that the intrinsic output of the diodes at 31.7
GHz varies by $\leq 1\%$.  At 14.5 GHz, instrumental effects resulted
in a variation of $\leq 2\%$ in the diode power output.  In the
Rayleigh-Jeans regime,
\begin{equation}
S_{\rm cal} = {2kT_{\rm cal}\over{\lambda^2}}\Omega_a,
\label{eq:stcal}
\end{equation}
so that the flux density of the diodes can also be determined by
measuring the diode antenna temperature.  These measurements are
performed in the standard manner by comparison with external
loads of known temperature and form the basis of our
calibration prior to 1996, when calibrator sources were not observed
on a regular basis.  The uncertainty in the diode temperature
determined from these measurements, when combined in quadrature with
the error in the non-linearity correction (see \S\ref{sec:rx}), is
$3.2\%$.  Note, however, that this calibration method requires a
separate measurement of
$\Omega_a$ from Eq.~\ref{eq:stcal}, so that uncertainties in the
temperature scale enter twice if $\Omega_a$ and $T_{\rm cal}$ are
not determined simultaneously.

At 31.7 GHz, the first calibration method yields
\begin{equation}
{\sigma^2_I/{I^2}} = (0.6\%)^2_{\Omega_m} + (3.3\%)^2_{S_{\rm cal}},
\end{equation}
for a total calibration uncertainty of $3.4\%$ from 1996, while the
second gives
\begin{equation}
{\sigma^2_I/{I^2}} = (0.6\%)^2_{\Omega_m} + (3.5\%)^2_{\Omega_a} +
(3.2\%)^2_{T_{\rm cal}},
\end{equation}
or a total calibration uncertainty of $4.7\%$ prior to 1996.  At 31.7
GHz, the mean of the calibration errors from the three independent
seasons (see below), weighted by the measurement error in the variance
from each season, gives a total calibration uncertainty of
\begin{equation}
{\sigma_{I}}_{31.7~\rm GHz} = 4.3\%.
\end{equation}

\section{Data Selection}
\label{sec:data}

On the 5.5-meter telescope, a total of three seasons of data were
obtained at 31.7 GHz from the winter of 1993 to the spring of 1996.  A typical
observing season at the OVRO lasts from early October until mid May.
On the 40-meter telescope, construction of the 14.5 GHz receiver was
completed in the spring of 1994, and only two seasons of data were
obtained.  Due to procedural differences in observing strategy between
the first and second halves of 1996, however, these data are divided
into two seasons which are analyzed separately below (for details, see
\cite{Leitchthesis}).

\subsection{Miscellaneous Edits}
\label{sec:misc}

The first level of FLUX editing consists of rejecting all data taken
when the receiver was saturated during periods of high atmospheric
water content.  Since the receiver typically remains saturated
during all four segments of the FLUX procedure (see \S\ref{sec:obs}),
these are readily identified as data with standard deviations
identically zero.  This rejects less than $1\%$ of the data.  Next, data
taken during excessively windy conditions, leading to tracking errors,
are identified by the excess time taken for a FLUX procedure to
complete.  Any data for which the difference between the actual and
expected duration is $> 1^s$ are excised, eliminating between $1-10\%$
of the data.  The range indicates the spread between
the three major divisions of data at each frequency, described above.

The non-linearity correction discussed in \S\ref{sec:rx} requires
interpolation of the total power onto the FLUX data.  Any FLUX for
which no bracketing power measurements are found within $1^h$ is
rejected, typically affecting $1\%$ or less of the data. No
non-linearity correction is applied to the 14.5 GHz data, as both CALs
and FLUXes are measured against the sky, and thus against the same
power background.

\subsection{Statistical Edits}
\label{sec:stat}

The noise diodes are sampled every 15 minutes to remove gain fluctuations from
the data.  The scatter in these diode measurements therefore provides
an unbiased
criterion for culling data affected by rapid gain
variations on both telescopes, and by atmospheric fluctuations on the 40-meter
telescope.  (Diodes at 14.5 GHz are measured directly
against the sky, while diodes on the 5.5-meter telescope are measured against
internal loads, reducing the contribution of atmospheric fluctuations
to scatter in the CALs by a factor of $\sim 100$.)  In this step,
$4-12\%$ of the data at 31.7 GHz and $14-40\%$ of the data at
14.5 GHz are rejected.

\begin{deluxetable}{lrrrrrr}
\footnotesize
\tablewidth{5in}
\tablenum{4}
\tablecaption{\label{tab4}\sc Data Edits}
\tablehead{\colhead{ } & \colhead{ } & \colhead{31.7 GHz} & \colhead{
}  & \colhead{} & \colhead{14.5 GHz} & \colhead{ }\nl\cline{2-4}\cline{5-7}\nl\colhead{Edit\tablenotemark{a}} & \colhead{1994} & \colhead{1995} & \colhead{1996} & \colhead{1995} & \colhead{1996\_1} & \colhead{1996\_2}}
\startdata
Saturated (\%)\dotfill & 0.05 &  0.77 & 0.15 & 0.20 & 0.64 & 0.03\nl
Outlier (\%)\dotfill & 0.46 &  0.24 & 0.14 & 0.21 & 0.15 & 0.37\nl
No Powers (\%)\dotfill & 0.69 &  1.41 & 0.00 & 0.00 & 0.00 & 0.00\nl
SD[2.5] (\%)\dotfill & 1.87 &  1.77 & 2.34 & 2.34 & 1.92 & 2.72\nl
sigSW[25,7.5] (\%)\dotfill & 2.25 &  1.67 & 2.07 & 0.02 & 0.00 & 0.00\nl
CALs (\%)\dotfill & 3.99 &  4.06 &11.39 &24.61 &39.80 &13.58\nl
Excess time (\%)\dotfill & 9.72 &  1.00 & 3.92 &24.60 & 1.29 & 3.67\nl
meanSD[25,2.0] (\%)\dotfill &22.64 & 36.87 &35.53 & 3.16 & 1.11 & 1.38\nl\nl
Total rejected (\%)\dotfill &41.67 & 47.78 &55.53 &55.13 &44.91 &21.75\nl\nl
Total\tablenotemark{b}\dotfill & 68,359  & 98,617 & 107,494 & 32,743 & 29,209
& 40,584\nl
\enddata
\tablenotetext{}{{\sc Note. ---} Percentage figures are percent of data points
that were rejected.}
\tablenotetext{a}{For an explanation of notation, see \S\ref{sec:data}.}
\tablenotetext{b}{Total number of data points before editing.}
\end{deluxetable}

Next, a series of statistical edits are applied to the data.  These
have previously been described in Myers et al. (1997), and the same
notation is retained here for consistency.  A combination of sliding
buffer edits is implemented, where in each case we form the test
statistic
\begin{equation}
t_i = X_i/\sigma_{i_{\it th}},
\end{equation}
where the data $X_i$ have been divided by $\sigma_{i_{\it th}}$, the
expected thermal noise for the $i^{th}$ measurement ($\sigma_{\rm rms}
= 1.4~{\rm mK\,s^{1/2}}$ at 31.7 GHz, and $\sigma_{\rm rms} = 
2.3~{\rm mK\,s^{1/2}}$ at 14.5 GHz).  The tested quantity $X$ can be the SD or
SW, a combination of the A-D integrations (see \S\ref{sec:obs}) which
cancels the signal in the far field, used to reject data during periods
of high residual atmospheric or instrumental fluctuations (see
\cite{myers}, \cite{Leitchthesis}).  Successive buffers of $N$ points
are constructed in time, where the width of the buffer is constrained
to be no more than two hours.  
For each buffer $j$, we compute the mean
\begin{equation}
\overline{t}_j = {1\over{N}}\sum^{j+N-1}_{i=j}t_i,
\end{equation}
and standard deviation
\begin{equation}
\sigma_j = \left[{1\over{N-1}}\sum^{j+N-1}_{i=j}(t_i -
\overline{t}_j)^2\right]^{1/2}.
\end{equation}
A point is rejected if there exists no buffer containing that point
for which either 
$\overline{t}_j < \overline{t}_{max}$ or $\sigma_j < \sigma_{max}$,
designated ``meanX[N, $\overline{t}_{max}$]'' and ``sigX[N, $\sigma_{max}$],''
respectively.  These filters are applied to the combined data from all
RING5M fields, as their primary purpose is to reject data affected by
the atmosphere, regardless of the field being observed.  In
addition to these buffer edits, we employ a simple point-by-point 
filter which rejects data for which $t_i > t_{\it max}$, designated 
``X[$t_{\it max}$].''  

A final edit, and the only edit in which the FLUX data themselves are
 used as a rejection criterion, is an iterative,
field-by-field $4\sigma$ outlier rejection.  This serves to reject isolated
spurious signals due to local radio-frequency interference, and
typically affects $< 0.5\%$ of the data.
The combined edits reject $50\%$ of the data at
31.7 GHz and $40\%$ at 14.5 GHz and are summarized in Table~\ref{tab4}.

The effect of the editing on the
weighted field means (see \S\ref{sec:fldmeans}) was investigated by
reducing the data for a wide
range of editing parameters; the mean standard deviation per field
introduced by varying the cutoffs is found to be 
$< 4$~\muK~(see Figure~\ref{fig5}).  

\bigskip
\centerline{\vbox{\epsfxsize=8.5cm\center{\epsfbox{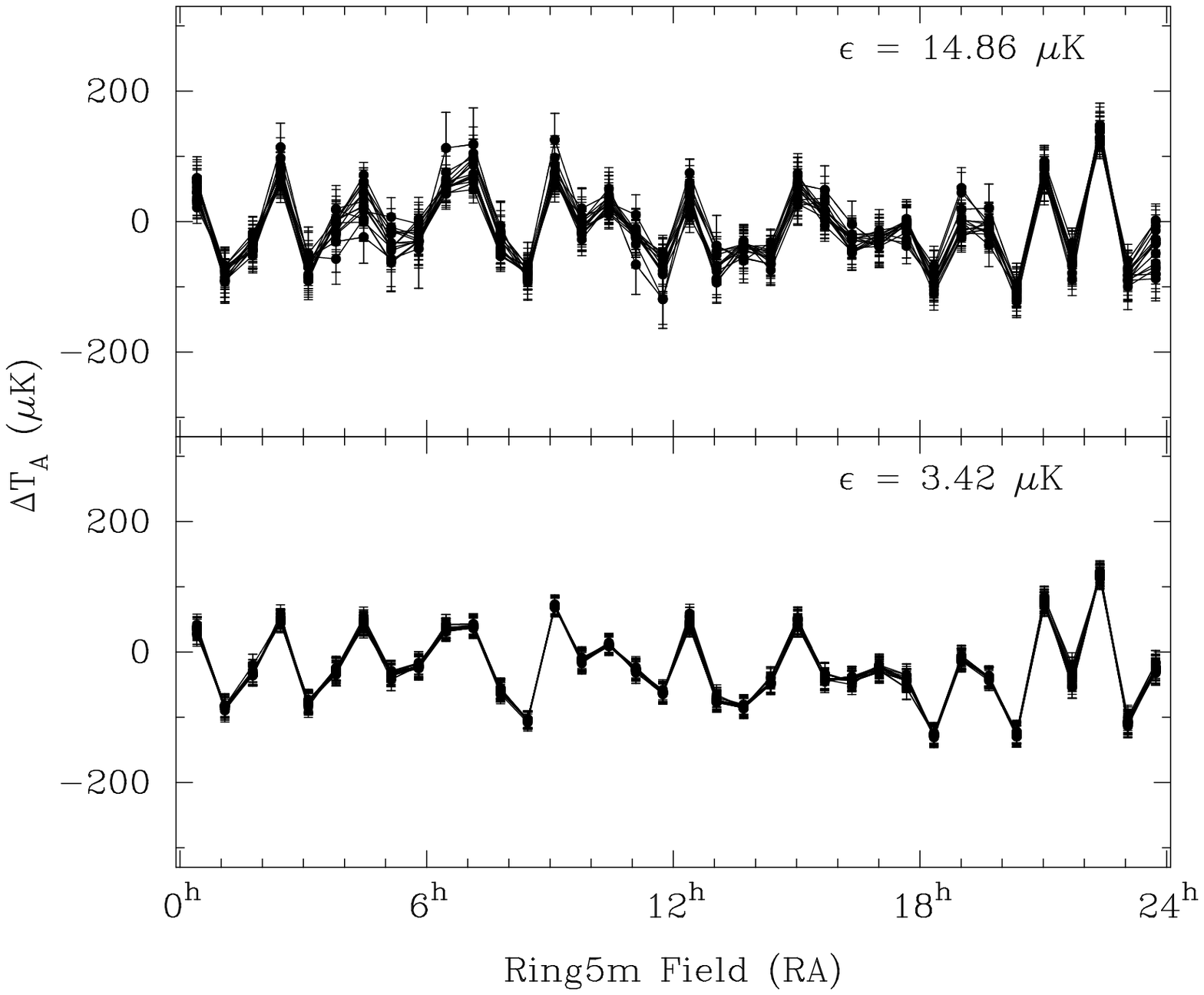}}
\figcaption{\footnotesize \label{fig5} Reduction of the 1994 RING5M
31.7 GHz data for 15 editing schemes for unweighted means (top panel),
and the same reductions using weighted means (bottom panel).  Even in
the case of the unweighted editing schemes, the mean standard
deviation per field (shown at the top right of each plot) is smaller
than the statistical error bars, and in the case of the weighted data,
is practically negligible.}}}
\bigskip

At 31.7 GHz, measurements of the noise diodes against the internal loads
can be used to estimate the residual contribution of the atmosphere to
the filtered data standard deviations; while the distribution of diode SDs is
consistent with the expected thermal noise, the distribution of FLUX
SDs against the sky
peaks at 1.4 times the thermal limit, with a considerable skew to
higher SDs.  We attribute this excess noise to residual atmospheric fluctuations
not removed by the fast switching (recall that on the 5.5-meter telescope, the
ANT and REF beams depart significantly beyond $\sim 0.5$~km, while the typical
scale height of water vapor is $\sim 2$~km).
At 14.5 GHz, the peak of the FLUX SD distribution is within 10\% of
the thermal limit, consistent with the lower atmospheric opacity at 14.5 GHz and
the better overlap of the 40-meter telescope beams.

\subsection{Field Means}
\label{sec:fldmeans}

Data for each RING5M field are acquired in 40-minute scans, during
which $\sim 21$ double-switched FLUXes are collected.  After the data
filtering described in \S\ref{sec:misc}-\S\ref{sec:stat}, the weighted
mean $\overline{\Delta T}_i \pm \epsilon_i$ for each field $i$ is computed as follows:
\begin{eqnarray}
\label{eq:wtmean}
\overline{\Delta T}_i &=& {1\over{W_1}}\sum^{N_i}_{j=1}{w_{\it ij}
\Delta T_{\it ij}}\\
\epsilon_i^2 &=& {W_2\over{W_1^2}}\sigma_i^2\\
\sigma_i^2 &=&
{{N_i}\over{{N_i}-1}}{1\over{W_2}}\sum^{N_i}_{j=1}{w_{\it ij}^2(\Delta
T_{\it ij} - \overline{\Delta T}_i)^2}\\
W_1 &=& \sum^{N_i}_{j=1}{w_{\it ij}}\\
W_2 &=& \sum^{N_i}_{j=1}{w^2_{\it ij}},
\end{eqnarray}
where $\Delta T_{\it ij}$ is the double-switched temperature from a single
FLUX procedure (see \S\ref{sec:obs}), and $N_i$ is the total number of
FLUXes recorded for field $i$.  Formally, the weights $w_{\it ij}$
should be chosen so that the sample mean is a maximum likelihood estimator for
the mean of the underlying distribution,
i.e., $w_{\it ij} = \sigma_{\it ij}^{-2}$.  In general, however, the standard
deviation $\sigma_{\it ij}$ reported with each datum (typically $1.4~\times$
the thermal noise at 31.7 GHz)
underestimates the scan standard deviation by a factor of 2-3,
presumably due to atmospheric fluctuations on timescales longer than a
single FLUX measurement.  The scan variance ${{\sigma^2_{\it ij}}_{\rm
sc}}$ is thus a better estimate of the real
error, and we take $w_{\it ij} = (\sigma^{2}_{\it ij} + 
{\sigma^{2}_{\it ij}}_{\rm sc})^{-1}$ when
computing statistics for the RING5M fields.

\bigskip
\centerline{\vbox{\epsfxsize=8.5cm\center{\epsfbox{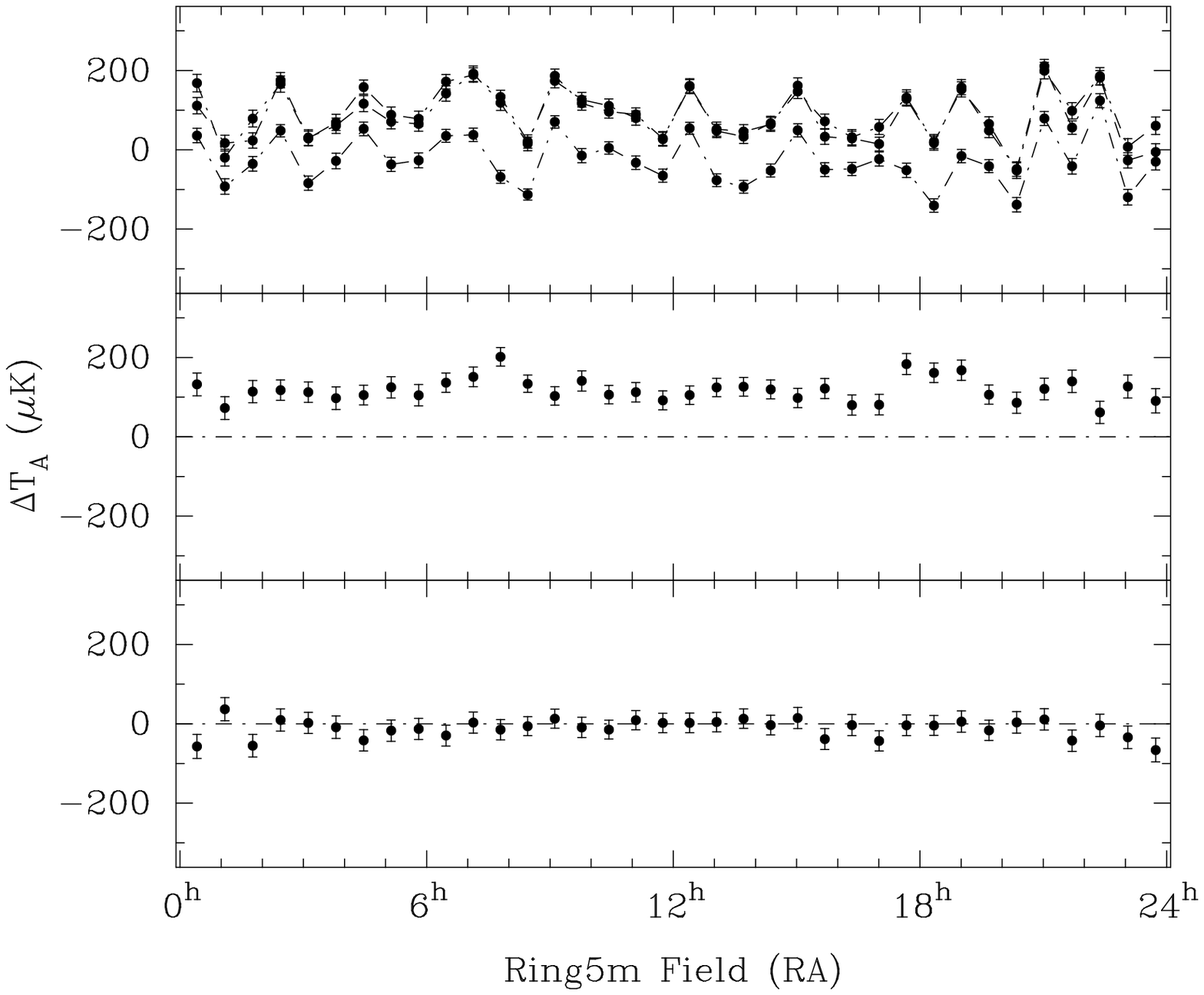}}
\figcaption{\footnotesize \label{fig6} RING5M results for 1994-1996.  (Top panel)
Three years of data, with no means subtracted.  Subtraction of the
1994 and 1995 data sets (middle panel) shows that the Ring has changed
by a constant offset ($\Delta T = 119.41~\mu$K, $\chi_r^2 = 1.34$),
while the 1996 and 1995 data sets are consistent with no change
($\Delta T = -9.93~\mu$K, $\chi_r^2 = 0.71$) (bottom
panel).}}}
\bigskip

\subsection{Mean Levels}
\label{sec:meanlevels}

The 31.7 GHz RING5M field means for 1994 -- 1996 are shown in
Figure~\ref{fig6}.  Between 1994 and 1995, the mean level of the
Ring changed by approximately
$120~\mu$K, while between 1995 and 1996, the difference is consistent
with zero.  Subtraction of the data sets shows that in each case the
shift is
consistent with an offset which is constant from field to field and
thus does not affect our estimate of the sky variance (a variable point
source contributes to the measured signal in the first field (see
\S\ref{sec:ptsrc}), which affects the two neighboring fields through
the double switching).  The constancy of these offsets implies that
they are instrumental in origin.

\bigskip
\centerline{\vbox{\epsfxsize=8.5cm\center{\epsfbox{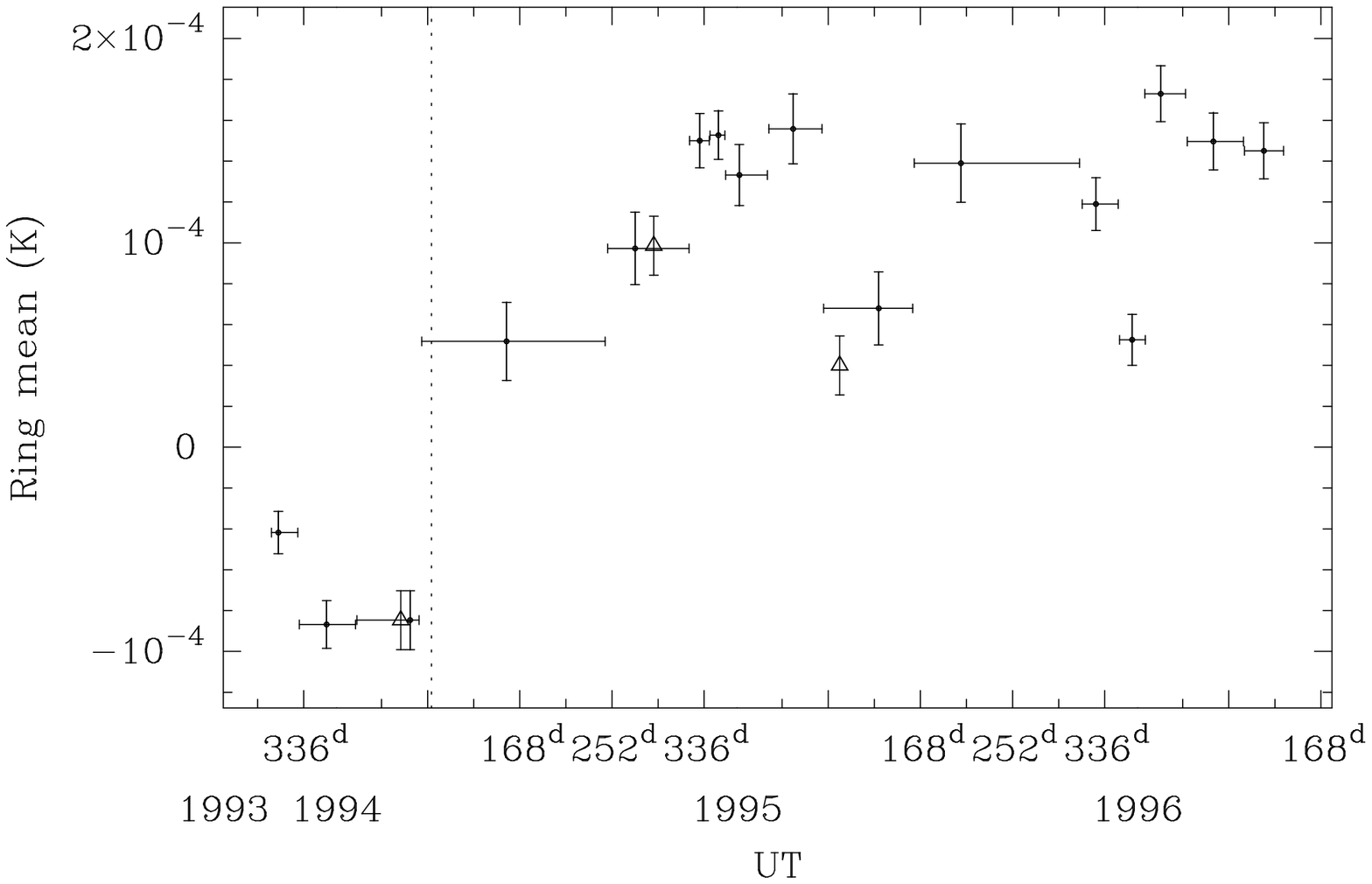}}
\figcaption{\footnotesize \label{fig7} Mean levels for the 31.7 GHz Ring data,
1994-1996, where bins contain equal numbers of data points.  The large step
in the mean level coincides with a HEMT change on day 91, 1994
(indicated by the dotted line).  Assuming variations in the mean
are given by Eq.~\protect\ref{eq:grad}, the predicted mean of the Ring at the
two epochs in 1994 and 1995 when measurements of the directional
isolations were made are indicated by open triangles, normalized to
the mean observed in 1993.}}}
\bigskip

Figure~\ref{fig7} shows that the largest shift in
the mean level occurred after a HEMT changeover on day 91 of 1994.
In \S\ref{sec:rx}, we showed that a mismatch
in the directional isolations of the Dicke switch can lead to incomplete
cancellation of ground or atmospheric temperature gradients.
Replacement of 
the HEMT will undoubtedly change the
impedance at the output port of the switch, quite likely resulting in a change
in isolation.  At 31.7 GHz, three measurements were made of the isolations, one
before the HEMT changeover, and two after.  We find that when the
error term given by Eq.~\ref{eq:grad} is normalized to the
RING5M mean in 1993, the mean levels predicted by the isolations
measured in 1994 and 1995 (shown as open triangles in Figure~\ref{fig7}) are in
excellent agreement with the data; in each case the predicted and
measured mean levels agree to within errors and imply a constant temperature
gradient of $\Delta\phi{\partial T\over{\partial\phi}} = -16~{\rm mK}$
across the beamthrow of the telescope.

\section{Concordance}
\label{sec:concord}

Edited and calibrated means for the 36 fields, representing a combined
total of approximately 4,500 hours of data,  are shown in
Figure~\ref{fig8}.  A mean level has been subtracted from each season.
Within each season, we test for internal consistency by dividing the data
set into two halves in time.  While the time spanned by a given season
is insufficient to permit a sensible comparison of data taken at night
to data taken during the day (see \S\ref{sec:data}), the data are
nevertheless quite repeatable on
timescales over which the field positions have precessed significantly
relative to the Sun; linear correlations (Pearson's $r$) between
halves fall in the range $r_{\rm \!obs} =
0.72-0.89$.  

At both frequencies, the data show structure well above
the noise and repeatable from year to year.  At 31.7 GHz, the
season-to-season deviations in the field means are Gaussian, and 
correlations between seasons are $r_{\rm \!obs} \geq
0.89$.  The probability of observing correlations this high under the
hypothesis that the sky temperatures are uncorrelated is $p(r >
r_{\rm\!obs}) \leq 6\times 10^{-8}$.  Under the hypothesis that the
data are completely correlated, as we should expect if the
signals are dominated by the microwave background, these correlations
fall within the
$68\%$ confidence region for $r_{\rm obs}$, given typical season-to-season
field errors of $\sim 20~\mu$K (antenna temperature).  Because of the double
switching, the RING5M field means are not statistically independent;
all probabilities quoted in this paper take the effect of the switching
into account.  In addition, we assume that the unswitched sky temperatures are
drawn from a Gaussian distribution.

\bigskip
\centerline{\vbox{\epsfxsize=8.5cm\center{\epsfbox{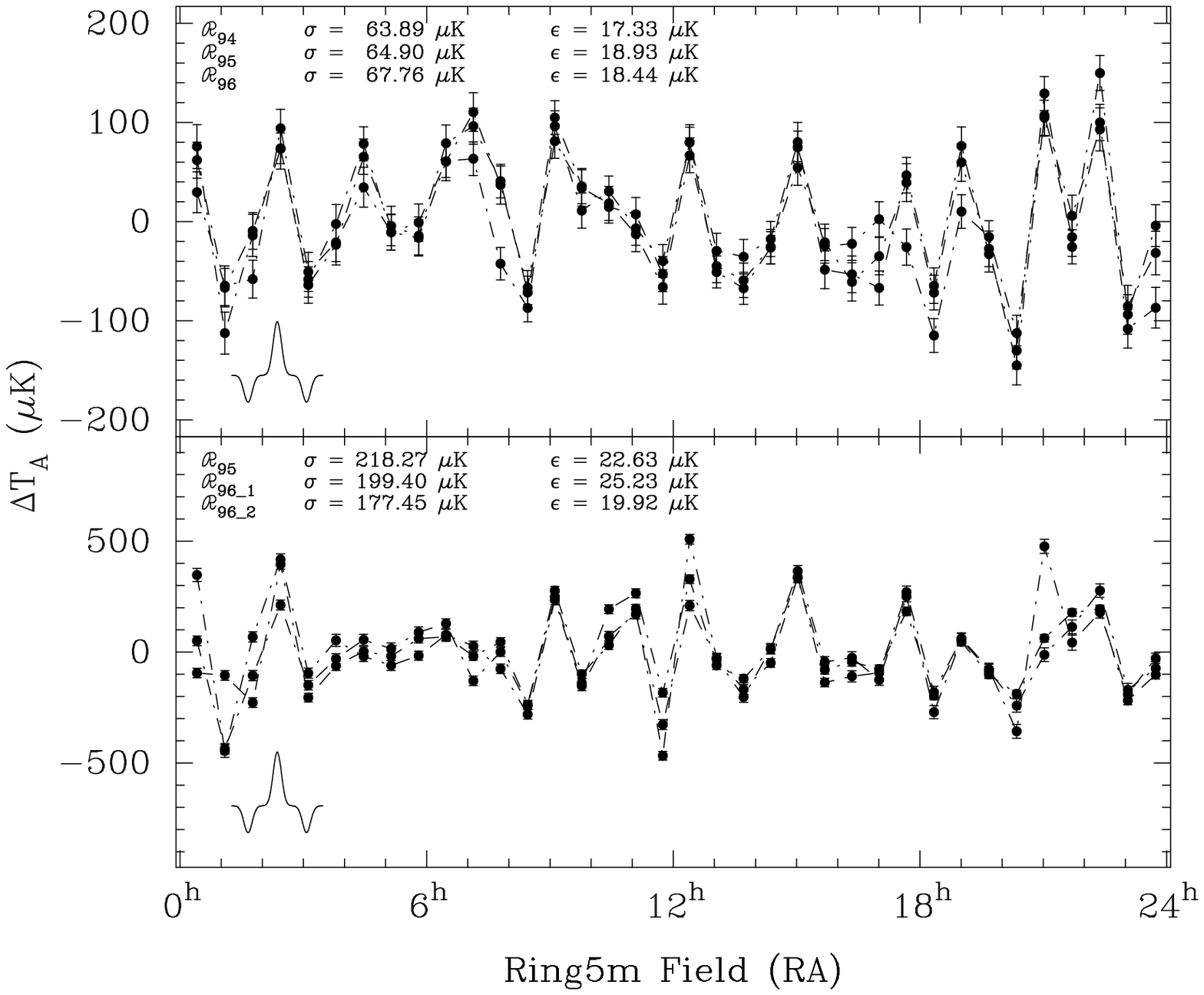}}
\figcaption{\footnotesize RING5M 31.7 GHz (top) and 14.5 GHz (bottom)
field means (antenna temperature) from winter of 1993 (${\cal
R}_{94}$) to winter of 1995 (${\cal R}_{96}$).  The {\rm rms} is
denoted $\sigma$, and $\epsilon$ is the mean error per field.  At
bottom left is the beam pattern produced by the double switching,
i.e., the effective point-spread function for the OVRO
telescopes.\label{fig8}}}}
\bigskip

At 14.5 GHz, the agreement is also good, with the exception of the
 RING5M field at $21^{\rm h}04^{\rm m}$
(OV5M2104), in which an anomalously high (and to date unexplained)
 signal was seen during the
first half of the 1995 season, and OV5M0024, which is dominated by a
bright variable point source (see \S\ref{sec:ptsrc}).  The two fields
adjacent to OV5M0024 are affected by the same point source due to the
double switching.  These effects collectively reduce season-to-season 
correlations at 14.5 GHz to the range $r_{\rm \!obs} = 0.72-0.87$.

\section{Terrestrial Contamination}
\label{sec:rfi}

Since the features we detect at both frequencies are fixed in sidereal
time, local radio frequency
interference (RFI) is an unlikely explanation for the structure observed
in the RING5M data.  Nonetheless, if there is ground-based
interference, the change in telescope elevation as a field is tracked
through transit
will introduce a characteristic parallactic angle dependence into the
data for any one field (see \S\ref{sec:dr21}), and we can look for
this signature in the RING5M data.  

The means for all 36 fields are shown in Figure~\ref{fig9}, binned in
parallactic angle.  We find no parallactic angle dependence in the
31.7 GHz data, indicating that these data are free from RFI
contamination.  The 14.5 GHz data, on the other hand, show a large
variation in amplitude with parallactic angle, indicating significant
contamination.  This pattern, however, occurs with the same amplitude
in each field (Figure~\ref{fig9} is a compilation of data from all 36
RING5M fields), demonstrating that RFI contributes only a mean level,
constant over long timescales, so that our measurement of the variance
should be unaffected.

\bigskip
\centerline{\vbox{\center{\hspace{0.25cm}\epsfxsize=4.0cm\epsfbox{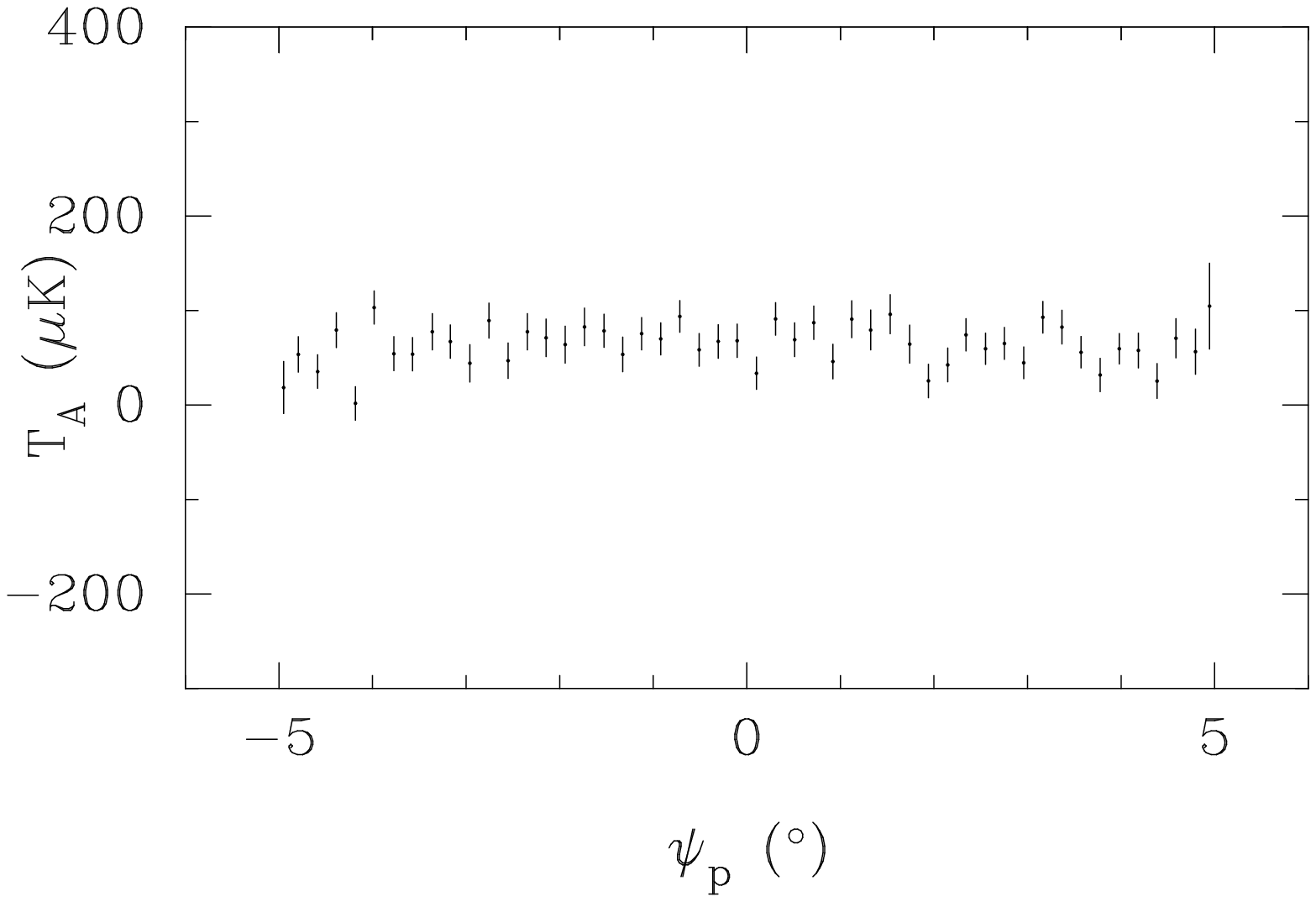}\hspace{0.25cm}\epsfxsize=4.0cm\epsfbox{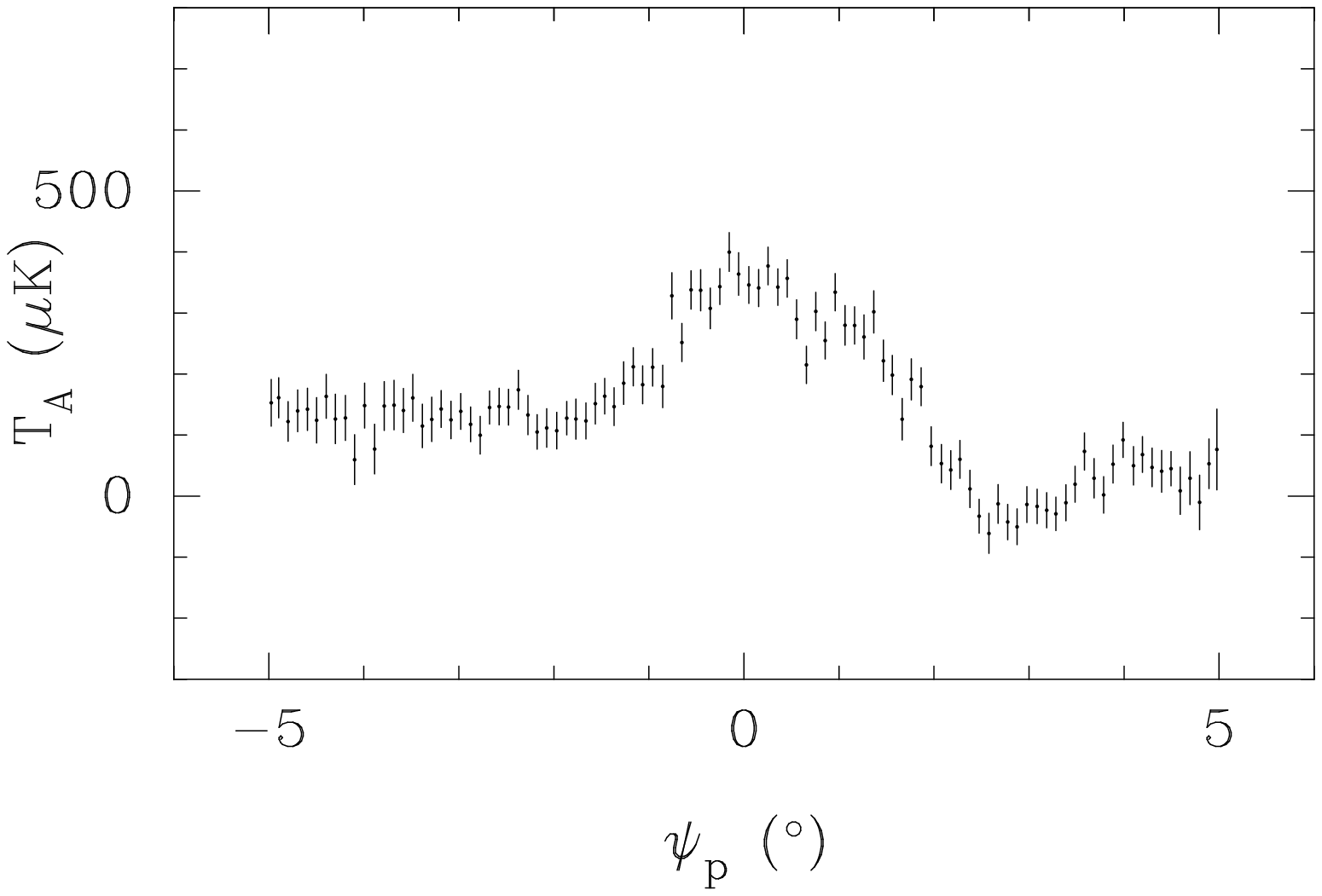}}
\figcaption{\footnotesize Co-added 31.7 GHz (top) and 14.5 GHz
(bottom) data from all 36 RING5M fields, binned in parallactic angle.
The 14.5 GHz data show the characteristic signature of interference
from the ground, while the 31.7 GHz data appear free of
contamination.\label{fig9}}}}
\bigskip

Although the parallactic angle dependence is a good indicator of the
robustness of our data to interference, we can definitively rule out
terrestrial contamination as a source of {\it structure} at 14.5 GHz
by observing the fields at lower culmination.  Since this changes the
position of the telescope beams only relative to the ground, signals
from the sky will remain unaltered, while any time-dependent
fluctuations from the ground will be shifted by 12 hours.  This test
was performed on 1996 Sep 16-30.  (The strength of the signals at 14.5
GHz permits a reasonable detection in this relatively short time.)
The comparison of the mean upper culmination 14.5 GHz data set with the lower
culmination data is shown in Figure~\ref{fig10}.  Field errors are
estimated by reducing the 1996 upper culmination data in 2-week
subsets; the error in the mean for each field is typically about
$62~\mu$K in antenna temperature.

\bigskip
\centerline{\vbox{\epsfxsize=8.5cm\center{\epsfbox{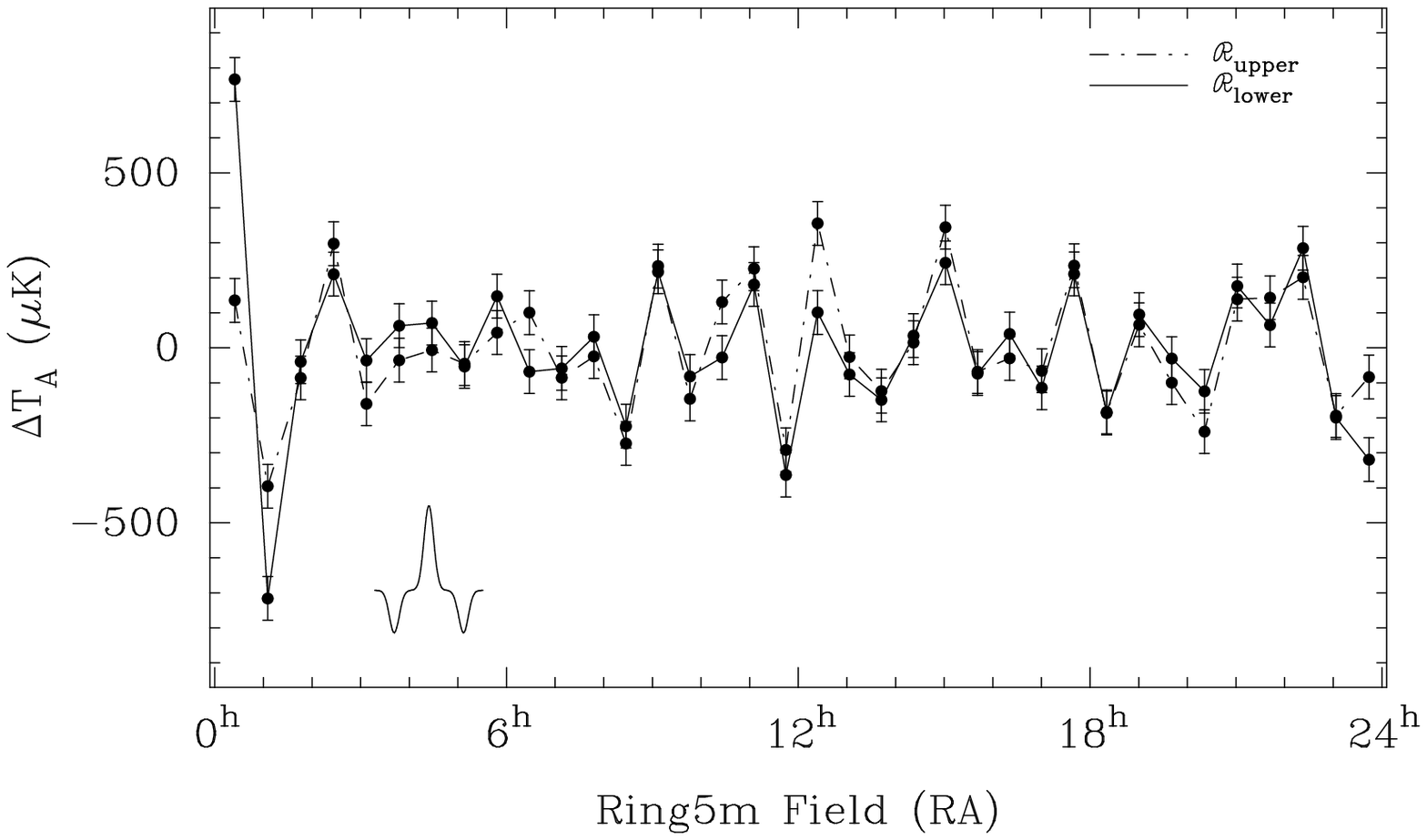}}
\figcaption{\label{fig10} \footnotesize 14.5 GHz upper culmination
field means (dot-dashed line), shown with the lower culmination data
(solid line).  The good agreement of these two data sets effectively
rules out terrestrial RFI as the source of the observed structure in
the Ring.  The large differences in the first, second and last fields
are due to discrete sources (see text).  At bottom left is the
effective beam pattern for double-switched
observations.}}}
\bigskip

The lower culmination data show the same structure, to within errors,
as the upper culmination data, demonstrating that the origin of these
signals is undoubtedly celestial.  The large differences in fields
{\sc OV5M0024}, {\sc OV5M0104} and {\sc OV5M2344} are due in part to
variability of a point source which dominates the signal in field {\sc
OV5M0024}, and to the slightly asymmetric 14.5 GHz double-switched
beam pattern, which is inverted at lower culmination relative to the
upper culmination beam.

\section{Discrete Sources}
\label{sec:ptsrc}

Mosaicked VLA observations covering the RING5M fields to the 3\% contour of
the 7\parcm{4} beam were made in 1994, with an rms sensitivity at 8.5
GHz of 0.21 mJy.  The sensitivity of the 5.5-meter telescope to point sources is
$4.1~{\rm mK}~{\rm Jy}^{-1}$, so that these observations allow
detection of any source contributing $\gtrsim 13$\muK~($4\sigma$) to
our highest frequency data with an accuracy of
$\lesssim3$\muK~($1\sigma$), 
assuming $\alpha \le +1$
(where $S_\nu\propto \nu^\alpha$).  During the period bracketing the 1996
RING5M season, we obtained multi-epoch VLA observations at 8.5 and 15 GHz of
the 39 sources (56 discrete components in all) found in the original 8.5
GHz survey.  These observations allow extrapolation of source flux
densities to 31.7 GHz and removal of variable contributions to the
RING5M data from sources which vary on timescales $\gtrsim 1$ month.

From October 1995 -- May 1996, RING5M sources were observed at 11 separate
epochs, in BnA, B, CnB, C and DnC configurations.  Due to the time
constraints of the 15 GHz observations (the approximate 10
minute sensitivity of the VLA at 15 GHz is 0.17 mJy, compared to 0.045
mJy at 8.5 GHz, so that 80\% of our observing time was spent at
15 GHz), not all sources could be observed at each epoch.  Sources 
which early-on showed significant variability were observed at each
epoch, while any remaining sources which could not be observed at a
given epoch were observed during the next.  Typical rms noise in the
maps was 0.25 mJy at 8.5 GHz and 0.5 mJy at 15 GHz.

\bigskip
\centerline{\vbox{\epsfxsize=7.5cm\center{\hspace{0.15cm}\epsfbox{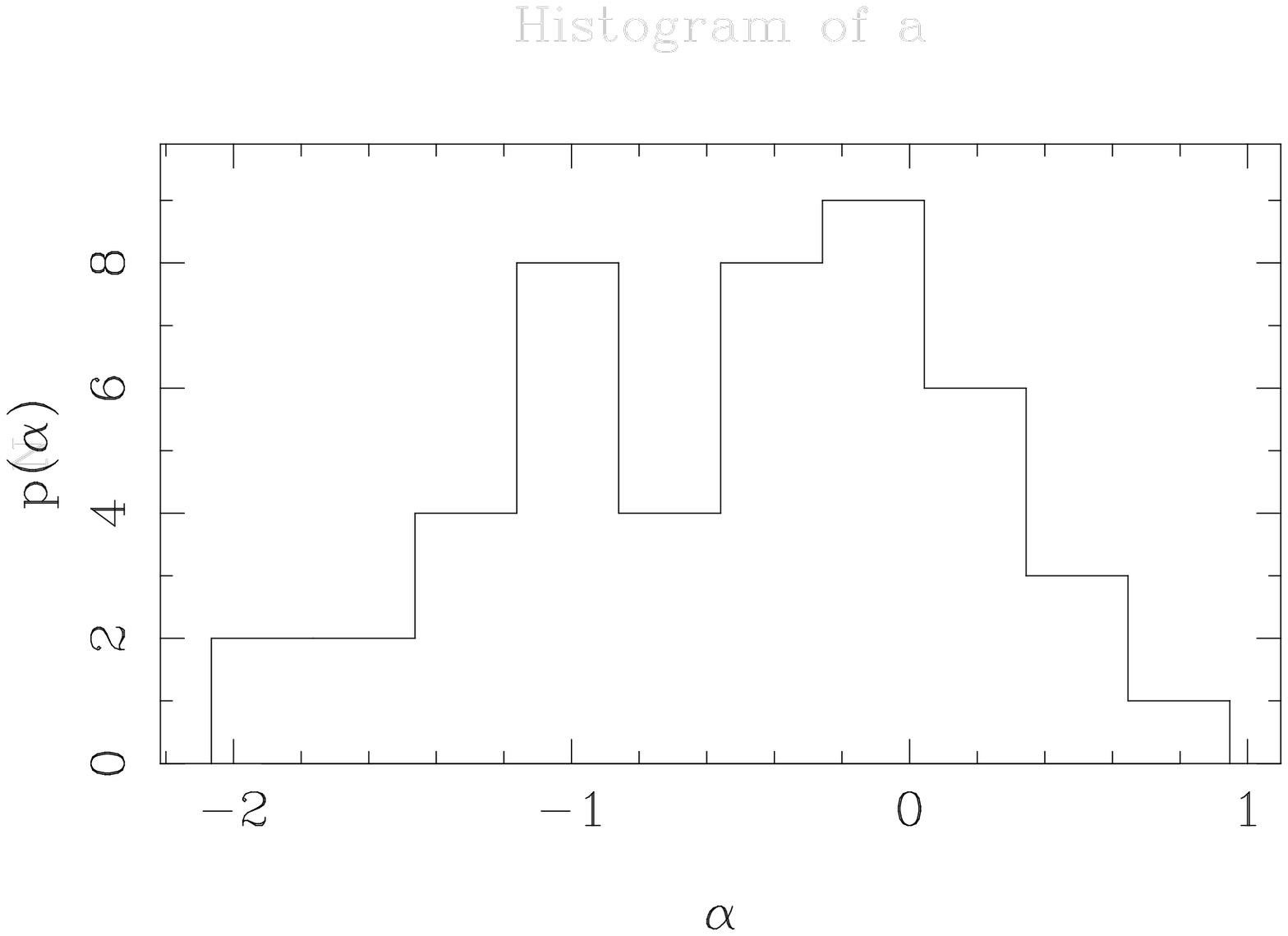}}
\figcaption{\footnotesize The histogram of spectral indices for
sources selected at 8.5 GHz, derived from the 47 discrete source
components detected at both 8.5 and 15 GHz.  Here, $p(\alpha)$ is the
number of sources with spectral index $\alpha\pm
0.15$.\label{fig11}}}}
\bigskip

For each source, the visibilities from all epochs at which the
source was observed are combined to form a single high-sensitivity map
at each frequency from which an accurate model for the source spatial
structure can be determined.  Several sources have multiple
components, consisting of a core and one or two prominent lobes, while
others show two point-like components closer than one would expect
from random superpositions of unrelated sources.  Model-fits to these
combined maps also provide a measurement of the mean flux density at each
frequency.  These are given in Table~\ref{tab5} for sources
detected at both frequencies, along with the source flux densities 
extrapolated to 31.7 GHz.  

Although sources detected at 8.5 GHz but not at 15 GHz --- nine in all ---
have falling spectra and will contribute negligibly to the RING5M
data, we can take a Bayesian approach (Sivia 1996) to
estimate their flux densities at higher frequencies.  The distribution
of spectral indices for sources selected at 8.5 GHz, shown in
Figure~\ref{fig11}, is constructed from the 47 components detected at
both 8.5 and 15 GHz.  Once the distribution of spectral indices is
known, the probability distribution of the 15 and 31.7 GHz flux
densities, given the measured flux density at 8.5 GHz and an upper
limit at 15 GHz,
can be constructed for each source via Monte Carlo simulation, yielding
the maximum likelihood estimates and corresponding $68\%$ confidence
intervals given in Table~\ref{tab5}.  

The estimated contribution of point sources to the 1996 31.7 and 14.5 GHz
data is shown in Figure~\ref{fig12}.
Few of the sources are bright enough or close enough to a
field center to contribute a significant signal, with the
notable exception of a $\sim100~$mJy source which dominates the signal in
field OV5M0024 and affects the two adjacent fields through the
double switching.  If we exclude OV5M0024 and
its flanking fields, the rms at 31.7 GHz due to discrete sources
is $12$\muK~(antenna temperature), or $< 4\%$ of the observed
variance, and $58$\muK~at 14.5 GHz, or $< 10\%$ of the variance
observed there.

\bigskip
\centerline{\vbox{\epsfxsize=8.5cm\center{\epsfbox{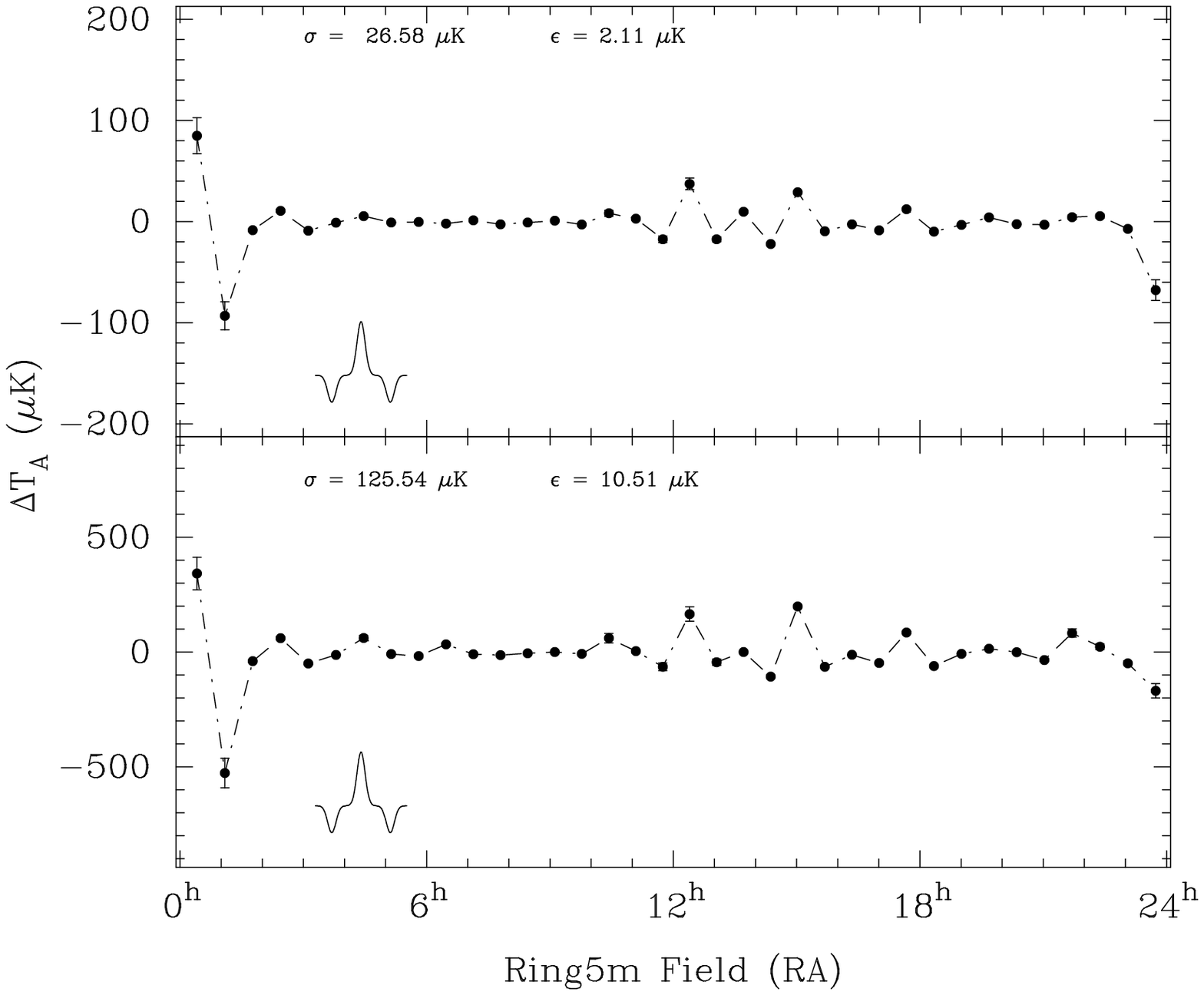}}
\figcaption{\footnotesize Point source contribution to the 31.7 GHz
Ring data (top), and to the 14.5 GHz data (bottom), extrapolated from
VLA monitoring at 8.5 and 15 GHz.  Error bars reflect uncertainties in
the source fluxes due to pointing only.  At bottom left is the
effective response to a point source near a Ring field
center.\label{fig12}}}}
\bigskip

Because sources were monitored during 1996, formal errors from point
source subtraction for this season are typically a few $\mu$K.  Since many of
the RING5M sources lie on the exponential cutoff of the $7\parcm{4}$
beams, however, small errors in pointing can produce large changes
in the antenna temperature produced by a source, so that the variance due to
pointing errors contributes significantly to the uncertainties.  Field errors in
Figure~\ref{fig12} are the computed standard deviations due to
pointing, assuming Gaussian azimuth and
zenith angle pointing errors with $\sigma_{A} = \sigma_{Z} = 0\parcm{3}$.
At 31.7 GHz, the largest uncertainty due to pointing is $17~\mu$K,
while at 14.5 GHz, the largest error is $70~\mu$K, comparable to the
variance of the brightest variable source in the Ring, making pointing the
dominant uncertainty.  We assume this to be true even for years when
the sources were not monitored.

In all subsequent analysis, a single combined data set is used at each
frequency.  Although the errors due to point-source
subtraction are not comparable from year to year, we can define a
mean error for each field 
\begin{equation}
\sigma = {\sum_j\sigma_{j}w_j\over{\sum_j w_j}},
\end{equation}
where $\sigma_{j}$ is the error from source subtraction from
season~$j$, and $w_j$ is the sum of the weights from season~$j$, i.e.,
\begin{equation}
w_{j} = \sum^{N_j}_k{1/\epsilon^2_{\it jk}},
\end{equation}
where $\epsilon_{\it jk}$ is the error in an individual FLUX measurement.
To these we add in quadrature the statistical errors for each field
mean and the error per field due to pointing uncertainties.  

The mean source-subtracted R-J temperatures $\Delta T_i$ and associated
uncertainties $\epsilon_i$ at each frequency are given in Table~\ref{tab6}.

\begin{deluxetable}{ccrrr}
\tablewidth{0pt}
\tablenum{5}
\scriptsize
\tablecaption{\label{tab5}\sc Summary of Source Flux Densities}
\tablehead{\colhead{$\alpha$ ({\rm J2000})} & \colhead{$\delta$ ({\rm J2000})} &
\colhead{$\rm{\overline{S}_{\rm 8.5~GHz}}$~(mJy)} &
\colhead{\phm{\tablenotemark{a}}$\rm{\overline{S}_{\rm 15~GHz}}$~(mJy)\tablenotemark{a}}
& \colhead{\phm{\tablenotemark{a}}$\rm{\overline{S}_{\rm 31.7~GHz}}$~(mJy)\tablenotemark{a}}}
\startdata
00:22:12.97 & 87:53:56.64 & $  2.10\pm 0.10$ & $0.78^{+0.16}_{-0.06}$ & $0.20^{+0.11}_{-0.04}$ \nl\hline
00:32:41.48 & 87:50:43.61 & $ \phm{\tablenotemark{b}}89.86\pm 0.56\tablenotemark{b}$ & $107.08\pm 0.49$       & $126.44\pm 0.60$       \nl\hline
02:22:09.58 & 87:52:18.73 & $  4.14\pm 0.10$ & $  3.67\pm 0.26$       & $3.12\pm 0.27$         \nl\hline
02:31:45.60 & 87:49:10.76 & $  4.49\pm 0.10$ & $  5.03\pm 0.34$       & $5.85\pm 0.47$         \nl\hline
04:28:27.47 & 87:58:32.65 & $ 17.89\pm 0.15$ & $  9.75\pm 0.31$       & $4.33\pm 0.15$         \nl\hline
06:30:29.55 & 87:53:18.41 & $  1.84\pm 0.11$ & $0.69^{+0.19}_{-0.06}$ & $0.18^{+0.14}_{-0.04}$ \nl      
06:30:29.55 & 87:53:19.64 & $  0.66\pm 0.11$ & $0.32^{+0.17}_{-0.10}$ & $0.09^{+0.22}_{-0.05}$ \nl\hline
06:34:14.18 & 87:56:23.83 & $  2.78\pm 0.11$ & $1.01^{+0.14}_{-0.06}$ & $0.26^{+0.09}_{-0.04}$ \nl\hline
07:02:43.09 & 87:45:09.00 & $\phm{\tablenotemark{b}}  7.67\pm 0.12\tablenotemark{b}$ & $  7.84\pm 0.40$       & $8.07\pm 0.46$         \nl\hline
08:56:31.32 & 87:47:58.19 & $  1.70\pm 0.11$ & $  2.10\pm 0.28$       & $2.79\pm 0.53$         \nl\hline
09:38:58.18 & 87:50:52.59 & $  2.23\pm 0.12$ & $0.84^{+0.13}_{-0.09}$ & $0.22^{+0.10}_{-0.05}$ \nl\hline
10:18:52.68 & 87:56:08.52 & $  8.32\pm 0.09$ & $  5.97\pm 0.42$       & $3.83\pm 0.27$         \nl      
10:20:12.52 & 87:56:11.76 & $  5.51\pm 0.09$ & $  4.87\pm 0.37$       & $4.13\pm 0.37$         \nl      
10:21:12.46 & 87:56:00.62 & $ 22.67\pm 0.09$ & $ 17.13\pm 0.41$       & $11.78\pm 0.27$        \nl\hline
10:56:50.10 & 87:47:46.16 & $  5.19\pm 0.10$ & $  5.97\pm 0.33$       & $7.76\pm 0.48$         \nl\hline
11:46:15.05 & 87:54:55.13 & $  4.88\pm 0.08$ & $  3.59\pm 0.29$       & $2.38\pm 0.24$         \nl      
11:46:13.76 & 87:54:56.52 & $  0.99\pm 0.08$ & $0.39^{+0.15}_{-0.05}$ & $0.11^{+0.12}_{-0.04}$ \nl\hline
11:48:37.64 & 87:42:05.72 & $\phm{\tablenotemark{b}} 13.13\pm 0.12\tablenotemark{b}$ & $  5.86\pm 0.30$       & $1.99\pm 0.12$         \nl      
11:48:39.63 & 87:42:09.74 & $\phm{\tablenotemark{b}}  2.62\pm 0.12\tablenotemark{b}$ & $  0.87\pm 0.30$       & $0.20\pm 0.12$         \nl\hline
11:53:24.39 & 87:56:06.25 & $  7.07\pm 0.10$ & $  3.94\pm 0.30$       & $1.80\pm 0.17$         \nl\hline
12:11:50.52 & 87:50:54.44 & $ 11.88\pm 0.32$ & $ 12.72\pm 0.31$       & $13.94\pm 0.39$        \nl\hline
12:16:17.38 & 87:51:24.27 & $\phm{\tablenotemark{b}} 20.45\pm 0.23\tablenotemark{b}$ & $ 21.36\pm 0.31$       & $18.76\pm 0.31$        \nl\hline
12:55:57.99 & 87:48:00.82 & $  3.72\pm 0.10$ & $  3.66\pm 0.27$       & $3.58\pm 0.33$         \nl\hline
13:00:16.65 & 87:45:09.98 & $  9.16\pm 0.11$ & $  5.10\pm 0.30$       & $2.33\pm 0.16$         \nl\hline
13:41:49.23 & 87:48:20.03 & $  1.89\pm 0.09$ & $  2.99\pm 0.29$       & $5.52\pm 0.70$         \nl\hline
14:27:16.54 & 87:47:39.89 & $\phm{\tablenotemark{b}}  4.70\pm 0.16\tablenotemark{b}$ & $  3.18\pm 0.26$       & $1.89\pm 0.20$         \nl\hline      
14:33:37.94 & 87:51:07.58 & $  2.91\pm 0.11$ & $1.06^{+0.11}_{-0.07}$ & $0.27^{+0.08}_{-0.04}$ \nl\hline
15:00:10.90 & 87:50:53.45 & $  6.17\pm 0.13$ & $  3.74\pm 0.61$       & $1.92\pm 0.38$         \nl      
15:00:10.12 & 87:50:57.62 & $  3.91\pm 0.13$ & $  3.30\pm 0.61$       & $2.63\pm 0.60$         \nl
14:59:50.43 & 87:50:07.46 & $  4.68\pm 0.13$ & $  4.21\pm 0.72$       & $3.65\pm 0.63$         \nl\hline
15:02:54.05 & 87:58:44.39 & $  3.07\pm 0.12$ & $  1.90\pm 0.38$       & $1.00\pm 0.28$         \nl\hline
15:11:29.40 & 87:55:43.30 & $  1.91\pm 0.11$ & $  2.29\pm 0.39$       & $2.92\pm 0.68$         \nl      
15:11:34.96 & 87:55:46.31 & $  1.24\pm 0.11$ & $0.49^{+0.21}_{-0.06}$ & $0.14^{+0.17}_{-0.05}$ \nl\hline
15:42:38.53 & 87:55:38.92 & $  4.39\pm 0.10$ & $  3.30\pm 0.31$       & $2.25\pm 0.26$         \nl\hline
17:03:23.72 & 87:45:10.68 & $  5.01\pm 0.09$ & $  4.61\pm 0.30$       & $4.12\pm 0.32$         \nl\hline
17:30:38.61 & 87:54:12.72 & $  6.12\pm 0.13$ & $  2.21\pm 0.25$       & $0.57\pm 0.09$         \nl
17:31:05.73 & 87:54:12.77 & $  2.46\pm 0.13$ & $  1.37\pm 0.25$       & $0.63\pm 0.18$         \nl
17:30:37.89 & 87:54:16.49 & $  4.85\pm 0.13$ & $  3.79\pm 0.26$       & $2.73\pm 0.22$         \nl\hline
17:31:04.27 & 87:56:33.13 & $  2.19\pm 0.24$ & $0.83^{+0.29}_{-0.10}$ & $0.23^{+0.21}_{-0.09}$ \nl\hline
17:39:48.20 & 87:49:52.79 & $  4.05\pm 0.08$ & $  3.99\pm 0.31$       & $3.91\pm 0.37$         \nl\hline
17:45:23.55 & 87:44:33.37 & $\phm{\tablenotemark{b}} 15.29\pm 0.13\tablenotemark{b}$ & $ 11.26\pm 0.28$       & $7.48\pm 0.20$         \nl\hline
19:00:58.12 & 88:01:38.76 & $\phm{\tablenotemark{b}}  6.80\pm 0.12\tablenotemark{b}$ & $  3.59\pm 0.36$       & $1.53\pm 0.19$         \nl\hline
19:29:34.27 & 87:55:02.19 & $  3.07\pm 0.09$ & $  3.96\pm 0.27$       & $5.56\pm 0.47$         \nl\hline
19:41:43.37 & 87:46:35.91 & $  2.55\pm 0.11$ & $  3.21\pm 0.38$       & $4.37\pm 0.66$         \nl\hline
20:32:28.60 & 87:59:24.91 & $ 13.96\pm 0.15$ & $  6.57\pm 0.28$       & $2.40\pm 0.12$         \nl
20:32:49.85 & 87:59:31.15 & $  3.79\pm 0.15$ & $  2.24\pm 0.28$       & $1.11\pm 0.19$         \nl
20:32:52.51 & 87:59:25.76 & $  2.87\pm 0.15$ & $  1.19\pm 0.28$       & $0.37\pm 0.14$         \nl
20:32:30.54 & 87:59:30.94 & $  2.38\pm 0.15$ & $  2.01\pm 0.28$       & $1.60\pm 0.32$         \nl\hline
21:41:01.80 & 87:58:10.62 & $  3.50\pm 0.07$ & $  1.45\pm 0.27$       & $0.45\pm 0.12$         \nl
21:41:53.75 & 87:57:51.41 & $  1.65\pm 0.07$ & $  1.54\pm 0.30$       & $1.40\pm 0.35$         \nl
21:41:52.32 & 87:57:53.02 & $  4.31\pm 0.07$ & $  1.98\pm 0.29$       & $0.70\pm 0.13$         \nl
21:41:03.45 & 87:58:10.02 & $  5.59\pm 0.07$ & $  3.80\pm 0.27$       & $2.27\pm 0.19$         \nl
21:41:41.22 & 87:57:55.77 & $  2.02\pm 0.07$ & $  0.99\pm 0.28$       & $0.38\pm 0.17$         \nl\hline
22:28:10.05 & 87:50:37.98 & $  9.02\pm 0.09$ & $  8.82\pm 0.30$       & $7.98\pm 0.31$         \nl\hline
23:16:20.13 & 87:49:41.46 & $  7.71\pm 0.99$ & $  3.97\pm 0.41$       & $1.63\pm 0.25$         \nl\hline
23:17:25.40 & 87:52:58.42 & $\phm{\tablenotemark{b}} 17.73\pm 0.15\tablenotemark{b}$ & $ 13.11\pm 0.30$       & $8.77\pm 0.22$         \nl
\enddata
\tablenotetext{a}{\parbox[t]{3.5in}{Asymmetric errors indicate a source was not
detected at 15 GHz and is a maximum likelihood estimate (see \S\ref{sec:ptsrc}).}}
\tablenotetext{b}{Variable source.}
\end{deluxetable} 			     				             

\begin{deluxetable}{lrrrr}
\tablewidth{0pt}
\tablenum{6}
\tablecaption{\label{tab6}\sc Source-Subtracted Ring Field Means\tablenotemark{a}}
\tablehead{\colhead{Field} & \colhead{$\overline{\Delta T}_{ 31.7}~$($\mu$K)} &\colhead{$\epsilon_{ 31.7}~$($\mu$K)\tablenotemark{b}} & \colhead{$\overline{\Delta T}_{14.5}~$($\mu$K)} & \colhead{$\epsilon_{ 14.5}~$($\mu$K)\tablenotemark{b}}}
\startdata
{\sc OV5M0024} & $-67.78$ & $36.88$ & $-710.82$ & $101.57$ \nl
{\sc OV5M0104} & $27.95$ & $32.78$ & $397.69$ & $96.38$ \nl
{\sc OV5M0144} & $-33.59$ & $17.15$ & $-94.52$ & $20.86$ \nl
{\sc OV5M0224} & $90.73$ & $15.94$ & $339.78$ & $21.75$ \nl
{\sc OV5M0304} & $-75.61$ & $16.40$ & $-168.44$ & $17.31$ \nl
{\sc OV5M0344} & $-21.54$ & $17.35$ & $-23.06$ & $18.68$ \nl
{\sc OV5M0424} & $79.59$ & $16.14$ & $-78.56$ & $24.04$ \nl
{\sc OV5M0504} & $-15.49$ & $16.32$ & $-56.43$ & $18.36$ \nl
{\sc OV5M0544} & $-13.89$ & $16.51$ & $47.66$ & $16.77$ \nl
{\sc OV5M0624} & $95.96$ & $15.75$ & $75.63$ & $18.99$ \nl
{\sc OV5M0704} & $126.83$ & $15.91$ & $-89.17$ & $17.21$ \nl
{\sc OV5M0744} & $13.87$ & $15.31$ & $-7.65$ & $17.91$ \nl
{\sc OV5M0824} & $-124.12$ & $13.88$ & $-357.49$ & $15.46$ \nl
{\sc OV5M0904} & $140.53$ & $14.57$ & $329.81$ & $15.61$ \nl
{\sc OV5M0944} & $46.24$ & $15.79$ & $-174.36$ & $16.63$ \nl
{\sc OV5M1024} & $19.74$ & $15.01$ & $86.55$ & $32.74$ \nl
{\sc OV5M1104} & $-7.82$ & $15.29$ & $262.51$ & $19.28$ \nl
{\sc OV5M1144} & $-56.77$ & $15.92$ & $-381.74$ & $27.99$ \nl
{\sc OV5M1224} & $51.82$ & $17.35$ & $340.51$ & $45.91$ \nl
{\sc OV5M1304} & $-40.51$ & $15.47$ & $-50.77$ & $25.38$ \nl
{\sc OV5M1344} & $-102.35$ & $15.40$ & $-208.90$ & $18.50$ \nl
{\sc OV5M1424} & $-8.84$ & $15.58$ & $107.09$ & $26.39$ \nl
{\sc OV5M1504} & $58.72$ & $16.95$ & $225.55$ & $35.13$ \nl
{\sc OV5M1544} & $-34.40$ & $16.27$ & $-65.48$ & $19.18$ \nl
{\sc OV5M1624} & $-67.51$ & $15.86$ & $-51.07$ & $18.61$ \nl
{\sc OV5M1704} & $-38.27$ & $15.76$ & $-61.29$ & $18.11$ \nl
{\sc OV5M1744} & $13.11$ & $16.33$ & $201.98$ & $20.53$ \nl
{\sc OV5M1824} & $-114.55$ & $15.50$ & $-170.47$ & $19.79$ \nl
{\sc OV5M1904} & $68.41$ & $16.03$ & $105.79$ & $17.35$ \nl
{\sc OV5M1944} & $-48.60$ & $15.31$ & $-121.67$ & $19.37$ \nl
{\sc OV5M2024} & $-191.67$ & $16.51$ & $-297.43$ & $19.70$ \nl
{\sc OV5M2104} & $176.57$ & $16.17$ & $266.81$ & $20.34$ \nl
{\sc OV5M2144} & $-21.59$ & $17.04$ & $152.51$ & $32.31$ \nl
{\sc OV5M2224} & $163.50$ & $16.90$ & $231.69$ & $24.88$ \nl
{\sc OV5M2304} & $-134.58$ & $17.65$ & $-207.03$ & $18.50$ \nl
{\sc OV5M2344} & $45.92$ & $27.35$ & $204.82$ & $47.56$ \nl
\enddata
\tablenotetext{a}{Equivalent R-J temperature.}
\tablenotetext{b}{Errors are $1\sigma$ rms in the sample mean.}
\end{deluxetable}

\bigskip
\centerline{\vbox{\epsfxsize=8.5cm\center{\epsfbox{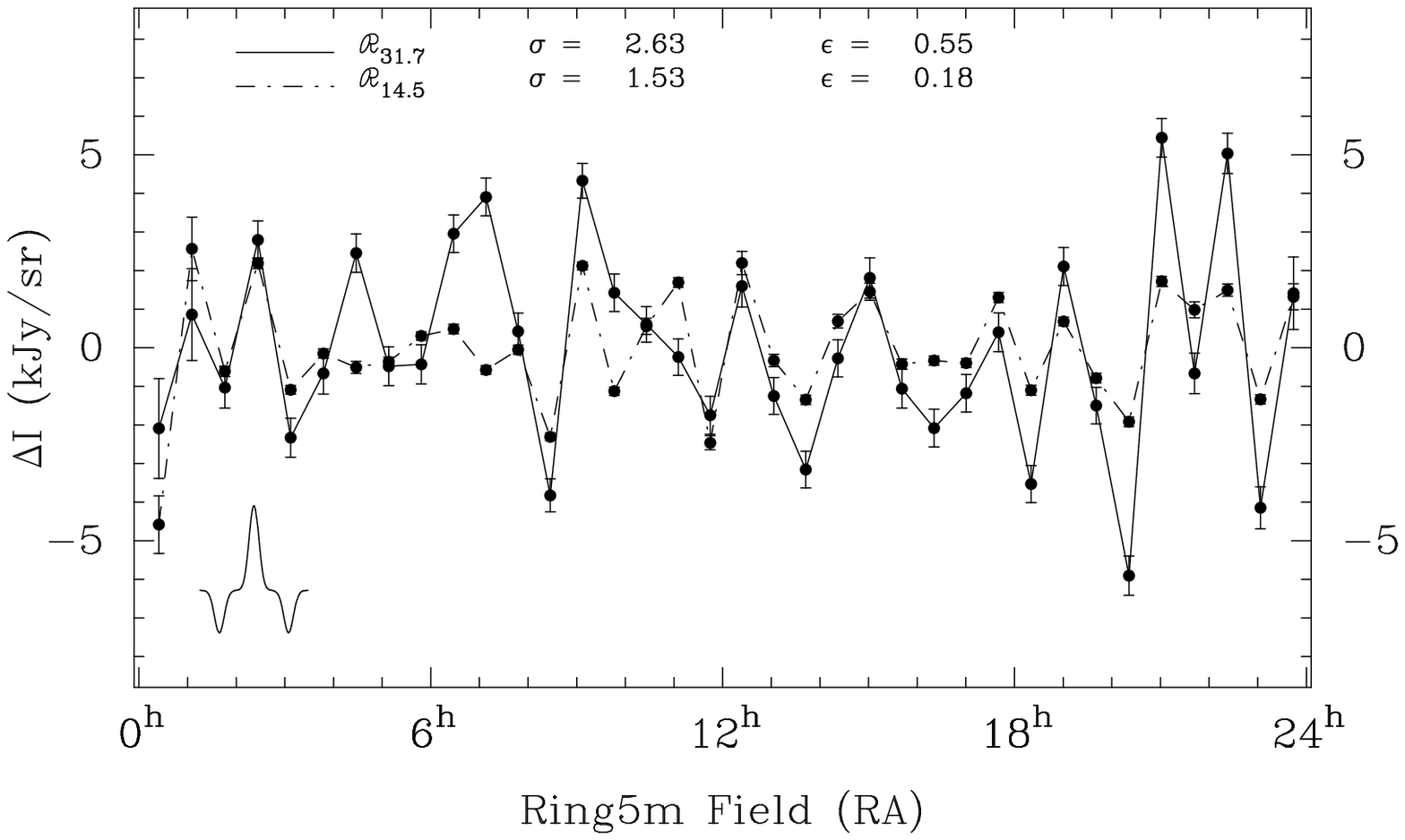}}
\figcaption{\footnotesize Mean source-subtracted 31.7 GHz (solid line)
and 14.5 GHz (dotted line) RING5M data, to equal intensity scales.
See Eq.~\protect\ref{eq:convert} for the conversion from $\Delta I$
(kJy/sr) to $\Delta T_B$ (K).\label{fig13}}}}
\bigskip

\section{Foregrounds}
\label{sec:foregrounds}

Mean (1993--1996) source-subtracted data sets at each frequency are shown in
Figure~\ref{fig13}.  The correlation between
frequencies is remarkably high, with $r_{\rm \!obs} = 0.81$ and
probability of observing a higher correlation than this under the
hypothesis that the data are
uncorrelated given by $p(r > r_{\rm \!obs}) = 2.7\times 10^{-6}$.  The
good agreement between several years of data from two independent
instruments demonstrates that the observed structure in the RING5M
data must
originate outside the telescopes.  Furthermore, as discussed in
\S\ref{sec:rfi}, observations of the RING5M fields at lower culmination
rule out interference from the ground.  We are therefore confident
that the signals are celestial in origin.

Since the CMBR specific intensity is approximately $I_{\rm cmb} =
2kT_{\rm cmb}\nu^2/c^2$, signals from the microwave background
should be reduced at 14.5 GHz by a factor of $\sim 5$ relative to
those at 31.7 GHz.  As can be seen in Figure~\ref{fig13},
however, many of the RING5M fields, notably in the regions $0^{\rm h}-3^{\rm h}$
and $12^{\rm h}-18^{\rm h}$, have equal intensities at the two
frequencies, suggesting that the emission mechanism may be thermal
bremsstrahlung.  On the other hand, the regions $4^{\rm h}-8^{\rm h}$
and $20^{\rm h}-23^{\rm h}$ show the spectral signature of the CMBR.

The nature of the steep-spectrum (in temperature) signals seen in the
RING5M data has been investigated by \cite{Leitch} (hereafter L1).
There we used a maximum likelihood test (described in detail in
\S\ref{sec:cmbr}) to rule out contamination by flat-spectrum
foregrounds under the
assumption that the CMBR is observed in the presence of a single
foreground.  We model the RING5M field means in the Rayleigh-Jeans
regime as
\begin{equation}
\Delta {T_i}_{\rm obs} = \Delta {T_i}_{\rm cmb} + \Delta {T_i}_{\rm fore},
\end{equation}
where $\Delta {T_i}_{\rm fore} \propto\nu^\beta$.
Given two frequencies $\nu_1$ and $\nu_2$, we can eliminate the
$\Delta {T_i}_{\rm fore}$ and solve for the CMBR component in each
field as a function of the unknown spectral
index $\beta$;
\begin{equation}
\Delta {T_i}_{\rm cmb}(\beta) = {\Delta {T_i}_{\rm obs}(\nu_1)\nu_1^{-\beta} -
\Delta {T_i}_{\rm obs}(\nu_2)\nu_2^{-\beta}\over{\alpha(\nu_1)\nu_1^{-\beta} -
\alpha(\nu_2)\nu_2^{-\beta}}},
\label{eq:sep}
\end{equation}  
where $\alpha(\nu)$ is a factor which corrects for the R-J
approximation to a true blackbody spectrum (at 31.7 GHz, $\alpha =
0.974$; see \cite{Leitchthesis}).
As can be seen in Figure~\ref{fig14}, foregrounds with
temperature spectral indices $\beta > -1.7$ can be
ruled out at the $3\sigma$ level. 

\bigskip
\centerline{\vbox{\epsfxsize=6cm\center{\hspace{0.25cm}\epsfbox{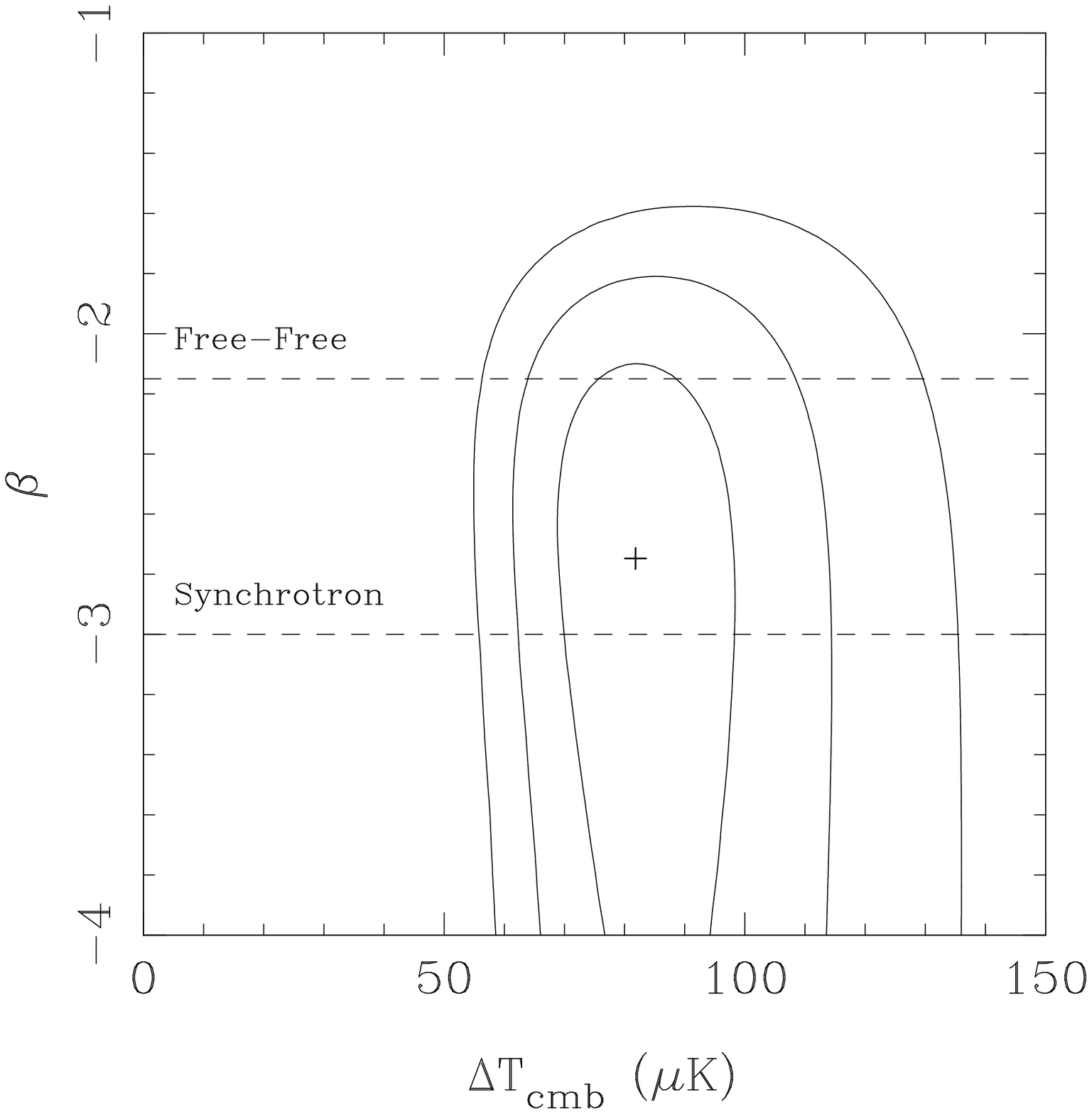}}
\figcaption{\footnotesize Likelihood function ${\cal L}(\Delta T_{\rm
cmb}, \beta)$ (see \S\protect\ref{sec:cmbr}) for the RING5M data,
assuming CMBR + single foreground of temperature spectral index
$\beta$.  Plotted are the $68\%$, $95\%$ and $99\%$ highest
probability density (HPD) intervals.\label{fig14}}}}
\bigskip

Although the RING5M data alone cannot provide much discrimination
among the steepest spectral indices in a two-component model, since
the CMBR component increasingly dominates for steeper foreground
spectral indices, we can use low-frequency maps of the NCP to
determine limits on the observed contribution of synchrotron emission
to the RING5M data.  At 325 MHz, maps from the WENSS survey
(\cite{wenss}) show no
detectable signals (see Figure~\ref{fig15}).  As discussed in L1,
we can use the rms in the 325 MHz map to rule out contamination of the
RING5M data by any foreground with temperature spectral index $\beta < -2.2$.

\bigskip
\centerline{\vbox{\epsfxsize=8.5cm\center{\epsfbox{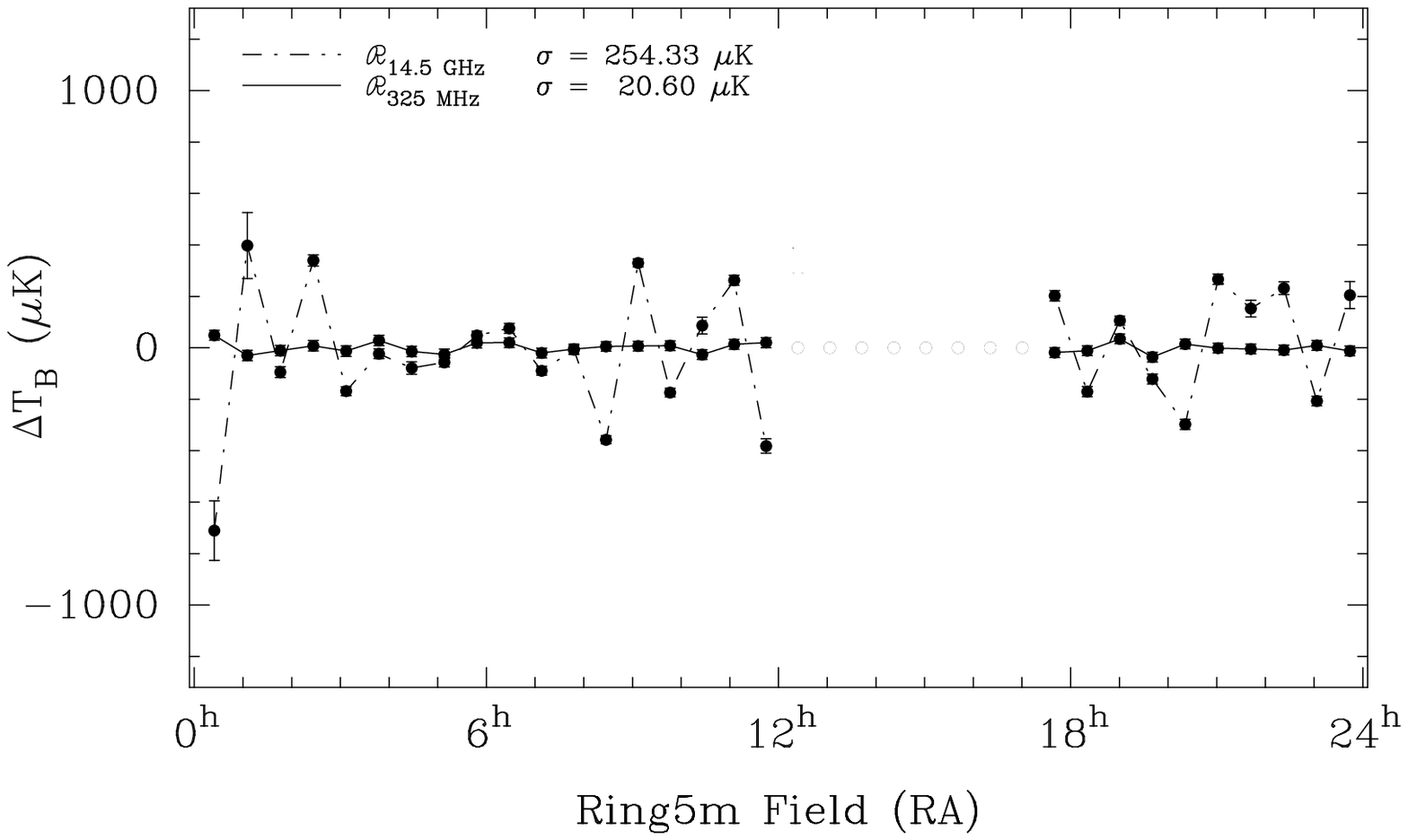}}
\figcaption{\footnotesize 14.5 GHz (dot-dashed line) and convolved 325
MHz map of the NCP (solid line) from the WENSS survey, extrapolated to
14.5 GHz assuming $\beta = -2.7$.  Standard deviations are quoted for
overlap region only (WENSS data are missing for fields {\sc OV5M1224 -
OV5M1744}).  The comparison demonstrates that the structure observed
at 14.5 GHz cannot be due to steep spectrum synchrotron
emission.\label{fig15}}}}
\bigskip

If the 14.5 GHz structure is due to thermal bremsstrahlung emission,
we would expect a considerable H$\alpha$ signature, unless the
temperature of the emitting gas is $T_e \gg
10^4~$K; lack of any detectable structure in H$\alpha$ maps
of the NCP in fact restricts the temperature of this component to
$\gtrsim 10^6$~K (see L1 for details).  In light of this,
flat-spectrum synchrotron or non-thermal emission from dust (see below)
may be a more likely explanation for these anomalous signals.

In L1, we also reported a significant correlation between the
steep-spectrum foreground at 14.5 GHz and {\it IRAS} $100~\mu$m
emission near the NCP --- independent confirmation that the structure
observed in the Ring is celestial in origin (see Figure~\ref{fig16}).
Draine and Lazarian (1998) have recently suggested that non-thermal emission
from spinning dust grains could produce emission with an apparent
free-free spectrum, while naturally accounting for the observed correlation.

Thermal emission from the dust itself is not expected to be a serious
contaminant at these frequencies (fits to the {\it COBE} DMR data
on $7^\circ$ scales suggest that $\sigma_{\rm\!dust} < 2$~\muK~at
31.4 GHz, while the DIRBE spatial template indicates that the power
spectrum of the dust is falling as ${\cal P}(l)\propto l^{-3}$ with
decreasing angular scale ($l\propto\theta^{-1}$); \cite{kogut}).

\begin{deluxetable}{lcc}
\tablewidth{5in}
\tablenum{7}
\tablecaption{\label{tab7}\sc Rms for Mean 1994-1996 31.7 and 14.5 GHz Data}
\tablehead{\colhead{}&\colhead{${\sigma_{31.7~\rm{GHz}}}$\tablenotemark{a}}&\colhead{${\sigma_{14.5~\rm{GHz}}}$\tablenotemark{a}}}
\startdata
Raw\dotfill                                    & $94.65~\mu$K & $249.93~\mu$K\nl
After point source subtraction\dotfill	            & $85.41~\mu$K & $238.06~\mu$K\nl
After foreground fitting, with $\beta_{\rm fore}=-2.2\dotfill$ & $82.51~\mu$K\tablenotemark{b} & $234.37~\mu$K\tablenotemark{c}\nl
\enddata
\tablenotetext{a}{R-J Temperature unless otherwise noted.}
\tablenotetext{b}{CMBR Brightness Temperature.}
\tablenotetext{c}{Extracted foreground component.}
\end{deluxetable}

\bigskip
\centerline{\vbox{\epsfxsize=8.5cm\center{\epsfbox{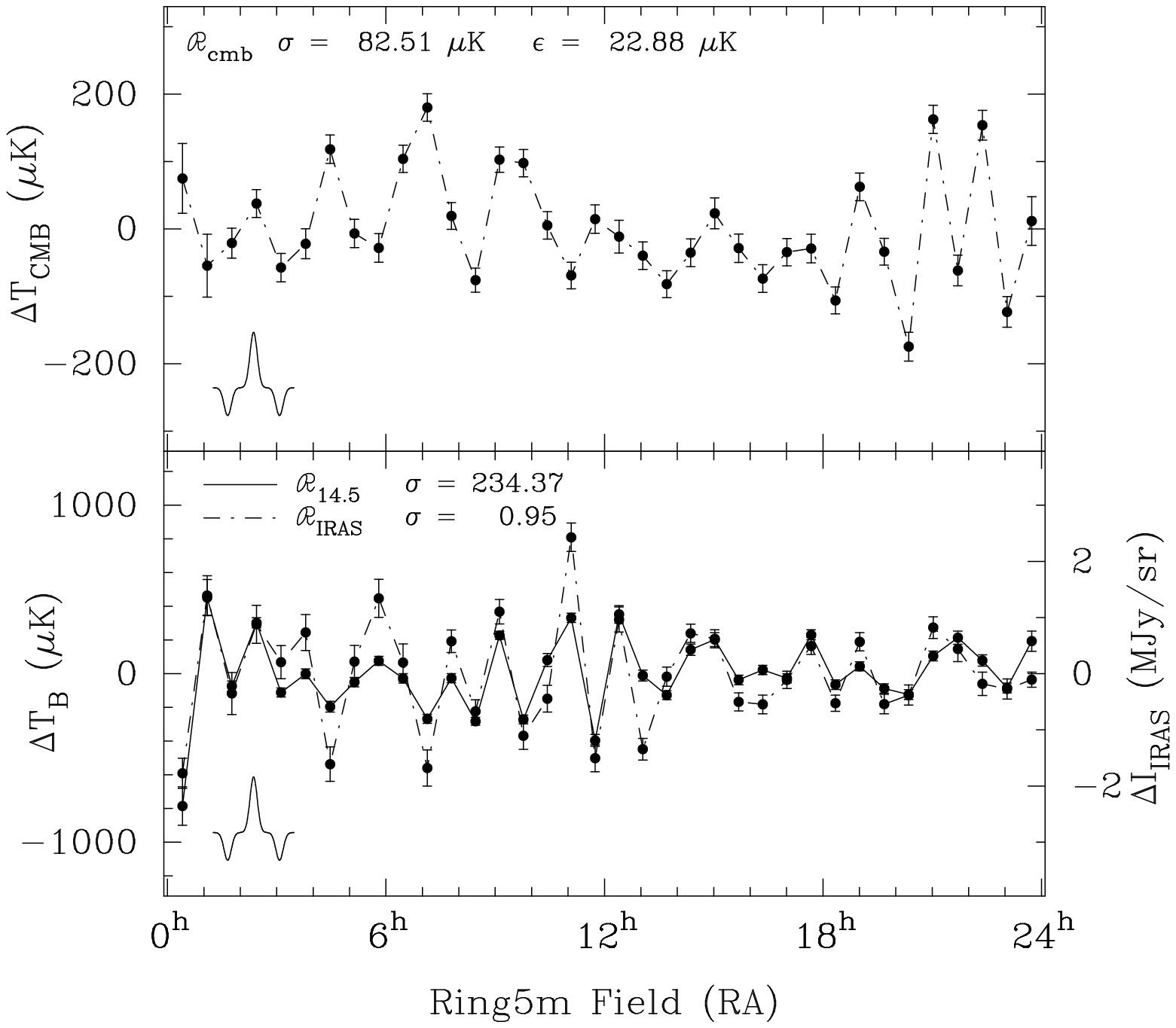}}
\figcaption{\footnotesize The extracted CMBR physical brightness
temperatures (at 31.7 GHz) (top panel), and the extracted foreground
equivalent brightness temperature (at 14.5 GHz) (bottom panel),
assuming a spectral index $\beta = -2.2$.  Note that at 14.5 GHz, even
the regions $3^{\rm h}-8^{\rm h}$ and $20^{\rm h}-23^{\rm h}$, in
which the CMBR dominates the combined signal (see
Figure~\protect\ref{fig13}), now show a noticeable correlation with
the {\it IRAS} $100~\mu$m intensities.\label{fig16}}}}
\bigskip

\section{Intrinsic Anisotropy}
\label{sec:cmbr}

In the foregoing discussion, we demonstrated that the signals detected in the
RING5M experiment are consistent with a combination of steep spectrum
($\beta \sim -2.2$) and blackbody emission ($\beta\sim 0$), the former
contributing $97\%$ of the variance at 14.5 GHz, the latter
responsible for $88\%$ of the variance at 31.7 GHz.
Low-frequency maps of the NCP were used to rule out contamination by
$\beta < -2.2$ emission, leaving us with a
foreground which either {\it is} free-free, or has very nearly the
same spectral dependence, in either case justifying the assumption of a
single foreground and allowing separate reconstruction of the CMBR
and foreground components from linear combinations of the 14.5 and
31.7 GHz data (see Figure~\ref{fig16}).  

Although the unexpected correlation of the 14.5 GHz data with Galactic
IR cirrus suggests that subtraction of a scaled IRAS template might be
a viable method of extracting the CMBR component, we feel that this
is not justified; CMBR plus a single foreground with the IRAS spatial 
template is a bad fit to our multifrequency data (best fit
$\chi^2_r\simeq 10$), as can also be seen by the large residual
differences in the lower panel of Figure~\ref{fig16}.
Furthermore, other methods which lead to more restrictive limits on
the CMBR amplitude make unjustifiable assumptions about the
distribution of the foreground.  We therefore take the extraction
outlined in the previous section to be the most conservative, as it
makes only the assumption that the frequency depedence of the
foreground can be modeled by a simple power law.

\addtocounter{footnote}{1}

The approach we take to estimating the CMBR variance is the
standard approach of Maximum Likelihood estimation (see, e.g.,
\cite{Sivia}).  We assume that the 36 field means are drawn from a
Gaussian distribution\footnote{Note that the Gaussian approximation is
equivalent to retaining the first three terms in the Taylor expansion
of any arbitrary distribution about its maximum.} on the sky, so that
the joint probability of the data set is given by
\begin{equation}
{\cal L} = {{{\exp(-{1 \over 2}{\bf t}^T{\bf C}^{-1}{\bf
t}})}\over{(2\pi)^{N/2}\left|{\bf C}\right|^{1/2}}},
\label{eq:like}
\end{equation}
where ${\bf t}$ is the data vector, with elements $\Delta T_i$ given
by Eq.~\ref{eq:sep}, and $\bf C$ is the associated $36\times 36$ covariance
matrix.  In general, the elements of ${\bf C}$ can be written
\begin{equation}
C_{\it ij} = \sigma^2_{\it ij} + {C_{\it ij}}_{\rm cmb},
\label{eq:varsky}
\end{equation}
where the $\sigma^2_{\it ij}$ are the temporal covariances from the
data (the diagonal elements $\sigma^2_{\it ii}$ are just the variances
of the field means).
The ${C_{\it ij}}_{\rm cmb}$ describe the predicted spatial variance
from the CMBR, and in general are given by the 2-pt correlation of the 
effective antenna pattern convolved with the sky temperature field 
(see \S\ref{sec:wfn}).

Once the likelihood function ${\cal L}(\Delta T_{\rm cmb}, \beta)$ is
constructed for the CMBR component, our estimate of the CMBR
variance $\Delta T^2_{\rm cmb}\equiv {C_{\it ii}}_{\rm cmb}$ is obtained by
maximizing $\cal L$ with respect to $\Delta T_{\rm cmb}$.  Implicit in
this construction is the
assumption of a spectral index $\beta$ for the anomalous foreground;
note however, that
even if all of the signal at 14.5 GHz were foreground, the relative
contribution of this foreground to the total variance at 31.7 GHz is
less than 25\% for spectral indices as flat as $\beta  = -2.2$ 
(see Table~\ref{tab7}).  Thus
our estimate of the CMBR variance is not strongly dependent on the
exact choice of foreground spectral index, as can also be seen from 
Figure~\ref{fig14}.

A conservative approach is simply to integrate out the dependence of
the likelihood on $\beta$, i.e.,
\begin{equation}
{\cal L}(\Delta T_{\rm cmb}) = \int{{\cal L}(\Delta T_{\rm cmb}, \beta)d\beta},
\label{eq:likeint}
\end{equation}
where we restrict the integration to bounds
defined by a reasonable prior, e.g., one uniform for $-3 < \beta < 2$,
since no Galactic foregrounds are known with $\beta < -3$ (see
\cite{RR}, who find that the low-frequency radio
spectral index near the NCP is $\sim -2.7$, and \cite{banday}, who
present evidence that the synchrotron spectral index steepens to
$\beta\sim -3$ at high frequencies).  The
integrated likelihood ${\cal L}(\Delta T_{\rm cmb})$ is shown in
Figure~\ref{fig17}.  The {\rm rms} obtained by maximizing ${\cal
L}(\Delta T_{\rm cmb})$ is 
\begin{equation}
{\delta T}_{\rm rms} = \dtrms,
\end{equation}
where quoted errors define the $68\%$ highest probability density
(HPD) confidence interval.

\bigskip
\centerline{\vbox{\epsfxsize=8.5cm\center{\epsfbox{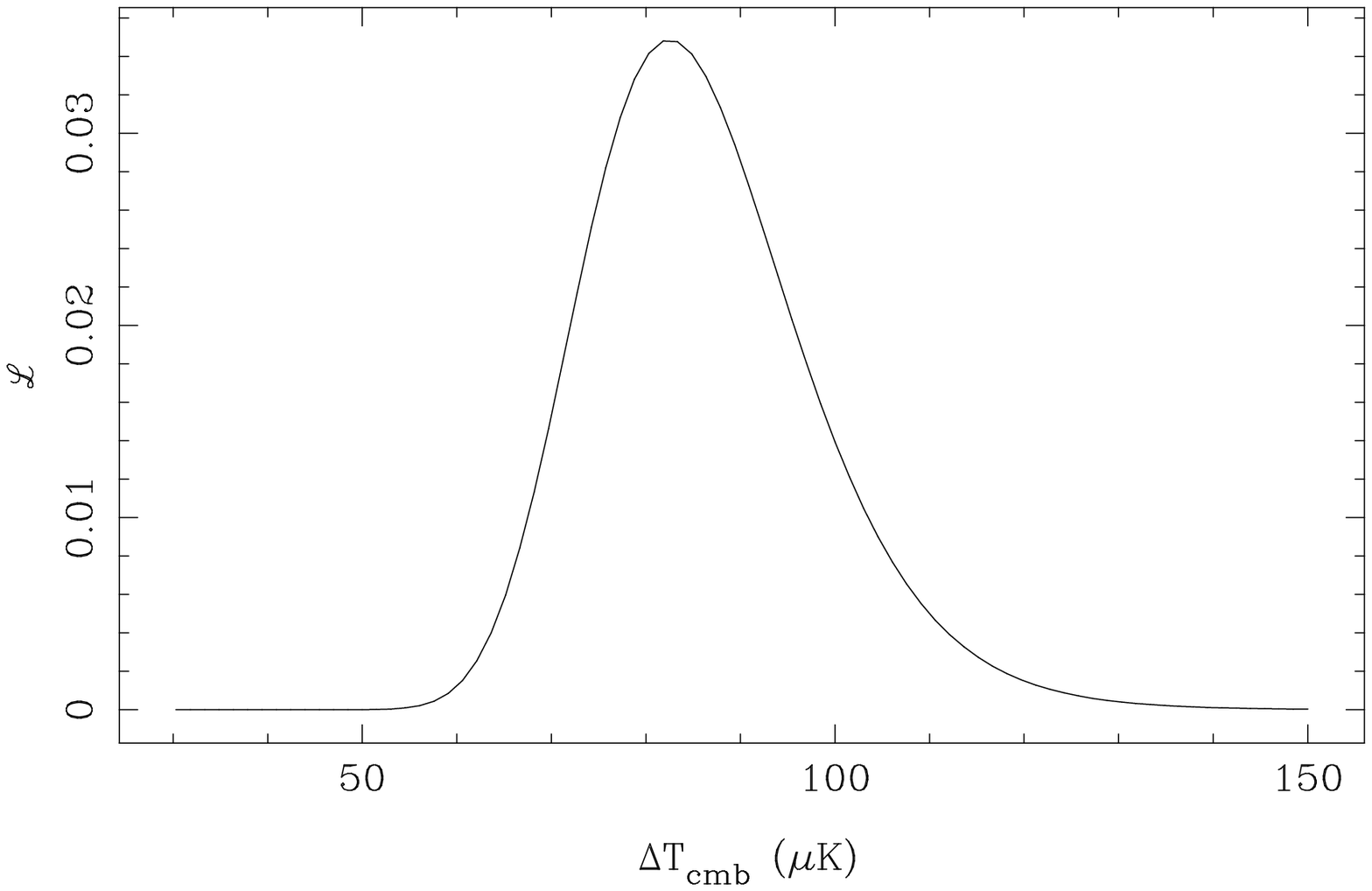}}
\figcaption{\footnotesize \label{fig17} Integrated likelihood function
${\cal L}(\Delta T_{\rm cmb})$ for the CMBR component.}}}
\bigskip

\section{Window Functions}
\label{sec:wfn}

The theoretical sky temperatures can be expanded in spherical
harmonics $T({\bf x}_i) = \sum_{l,m}a^m_l Y_l^m({\bf x}_i)$,
so that assuming rotational symmetry, the expected value of the 2-pt 
correlation for fields $i$ and $j$ separated by an angle
$\chi_{\it ij}$ on
the sky is given by
\begin{equation}
\langle{T_{i}\cdot T_{j}}\rangle =
{1\over{4\pi}}\sum_l(2l+1)C_l P_l(\cos\chi_{\it ij}),
\label{eq:tt}
\end{equation}
where $P_l$ are the associated Legendre polynomials, and $C_l =
\langle|a_l^m|^2\rangle$ are the elements of the theoretical angular
power spectrum (\cite{Peebles}).
What we measure are the theoretical sky temperatures
convolved with the telescope beam and switching strategy,
i.e., $\Delta T_i = T_i*B_i$ (for the RING5M experiment, $B_i(\theta,\phi)$ is
given by the function in Figure~\ref{fig1}), so that defining
${C_{\it ij}}_B \equiv B_i\otimes B_j$, where $\otimes$ denotes
cross-correlation, and letting $*$ denote
convolution, Eq.~\ref{eq:tt} gives 
\begin{eqnarray}
\nonumber
{C_{\it ij}}_{\rm cmb} &=& {1\over{4\pi}}\sum_l
(2l+1)C_l P_l(\cos\chi_{\it ij})*{C_{\it ij}}_B\\
&\equiv& {1\over{4\pi}}\sum_l
(2l+1)C_l W^{\it ij}_l
\label{eq:tobs2}
\end{eqnarray}
for the theoretical elements of the covariance matrix in
Eq.~\ref{eq:varsky}.  In CMBR parlance, the function
$W^{\it ij}_l \equiv P_l(\cos\chi_{\it ij})*{C_{\it ij}}_B$ is known as a
{\it window function}.  For the RING5M experiment, a good
approximation to the zero-lag ($\chi_{\it ii} \equiv 0$) window
function is given by 
$W^{\it ii}_l = W^{\it bm}_l\cdot W^{\it sw}_l$, where
$W^{\it bm}_l =
\exp[-l(l+1)\sigma^2]$ (\cite{silk}, \cite{bond1}), with $\sigma = 
3\parcm{13}$, and $W^{\it sw}_l = {3\over 2} - 2P_l(\cos{\chi_s})
+{1\over 2} P_l(\cos{2\chi_s})$, with $\chi_s = 22\parcm{16}$.

\bigskip
\centerline{\vbox{\epsfxsize=6cm\center{\hspace{0.25cm}\epsfbox{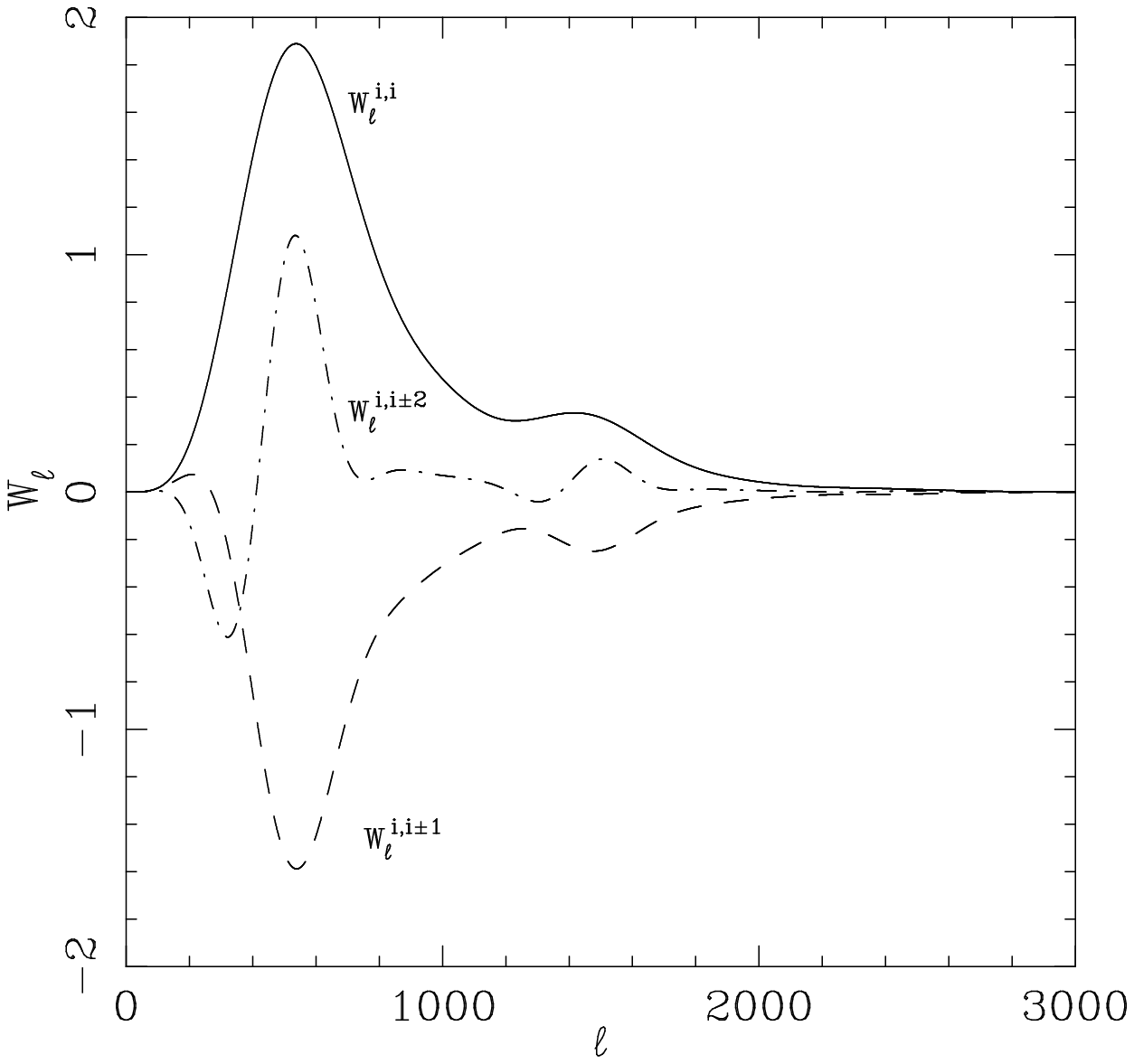}}
\figcaption{\footnotesize Comparison of $0$\arcm~($W^{i,i}_l$),
$22$\arcm~($W^{i,i+1}_l$) and $44$\arcm~lag ($W^{i,i+2}_l$) window
functions for the RING5M experiment, characterizing correlations
between neighboring fields.\label{fig18}}}}
\bigskip

\section{Data Correlations}
\subsection{Point Sources}
\label{sec:ptsrc_corr}

In general, point sources are not measured coincidentally with the
Ring data, but are instead assumed to have constant flux densities and
associated errors over timescales of a month (the typical time between VLA
flux monitoring sessions during 1996) or longer.  Since any source
affects at least three fields through the double switching,
subtraction of source contributions introduces correlated noise
between neighboring fields, contributing
\begin{equation}
\sigma^2_{\it ij} = \sum_k{\langle{b_{\it ik}}\rangle\langle{b_{\it
jk}}\rangle\sigma^2_{k}}
\end{equation}
to the covariance matrix, where the $b_{\it ik}$ are the beam
weighting factors for the $k^{\it th}$ source in the $i^{\it th}$
field, and $\sigma_k$ is the error associated with the flux density of
the $k^{\it th}$ source.  Although these covariances are included in
the likelihood analysis, the effect is negligible, even for the
1994-1995 RING5M data, for which source errors are enlarged to include
variability.

\subsection{Noise Correlations}
\label{sec:noise_corr}

An analysis of the long-term noise characteristics of the Ring5m data
indicates the presence at 31.7 GHz of correlated noise between fields
separated by $22\arcm$.  The amplitude of this correlated noise is 
approximately $40$\muK, or  $3-4\%$ of the uncorrelated noise
in a single scan on a Ring field (see \S\ref{sec:fldmeans}).  Analysis
of subsets of the 31.7 GHz data suggest no obvious source for this
component; we note however that its amplitude is of the same order as
the observed season-to-season fluctuations in the 31.7 GHz mean levels
(see \S\ref{sec:meanlevels}), suggesting that the same component may
be responsible for both.

While the origin of this correlated noise is not well understood,
independent confirmation of its presence can be seen through its
effect on the data correlations; the mean Ring5m data set shows only
half of the anti-correlation for nearest-neighbor fields expected for a 3-beam
chopped experiment, $C_{i,i\pm1}/C_{ii} = -2/3$.  

Our approach is to include the noise correlation as a free parameter
in our model for the covariance matrix, i.e., we let
\begin{equation}
\sigma^2_{\it ij} = \delta_{\it ij}\sigma^2_{\it ii} + \delta_{\it
i\pm1,j}\sigma^2_{\rm n}
\end{equation}
in Eq.~\ref{eq:varsky}, where $\sigma_{\rm n}$ is the amplitude of the
correlated noise, and $\sigma^2_{\it ii}$ are the variances in the
field means, as before.
As with the foreground spectral index, we integrate out the dependence
on the noise correlation amplitude to obtain an estimate of the CMBR variance.
Thus, the full form of the likelihood function ${\cal L}(\Delta T_{\rm cmb})$ in
Eq.~\ref{eq:likeint} is 
\begin{equation}
{\cal L}(\Delta T_{\rm cmb}) = \int\!\!\!\int{{\cal L}(\Delta T_{\rm
cmb}, \beta, \sigma_{\rm n})d\beta\,d\sigma_{\rm n}}.
\end{equation}

\bigskip
\centerline{\vbox{\epsfxsize=6cm\center{\hspace{0.15cm}\epsfbox{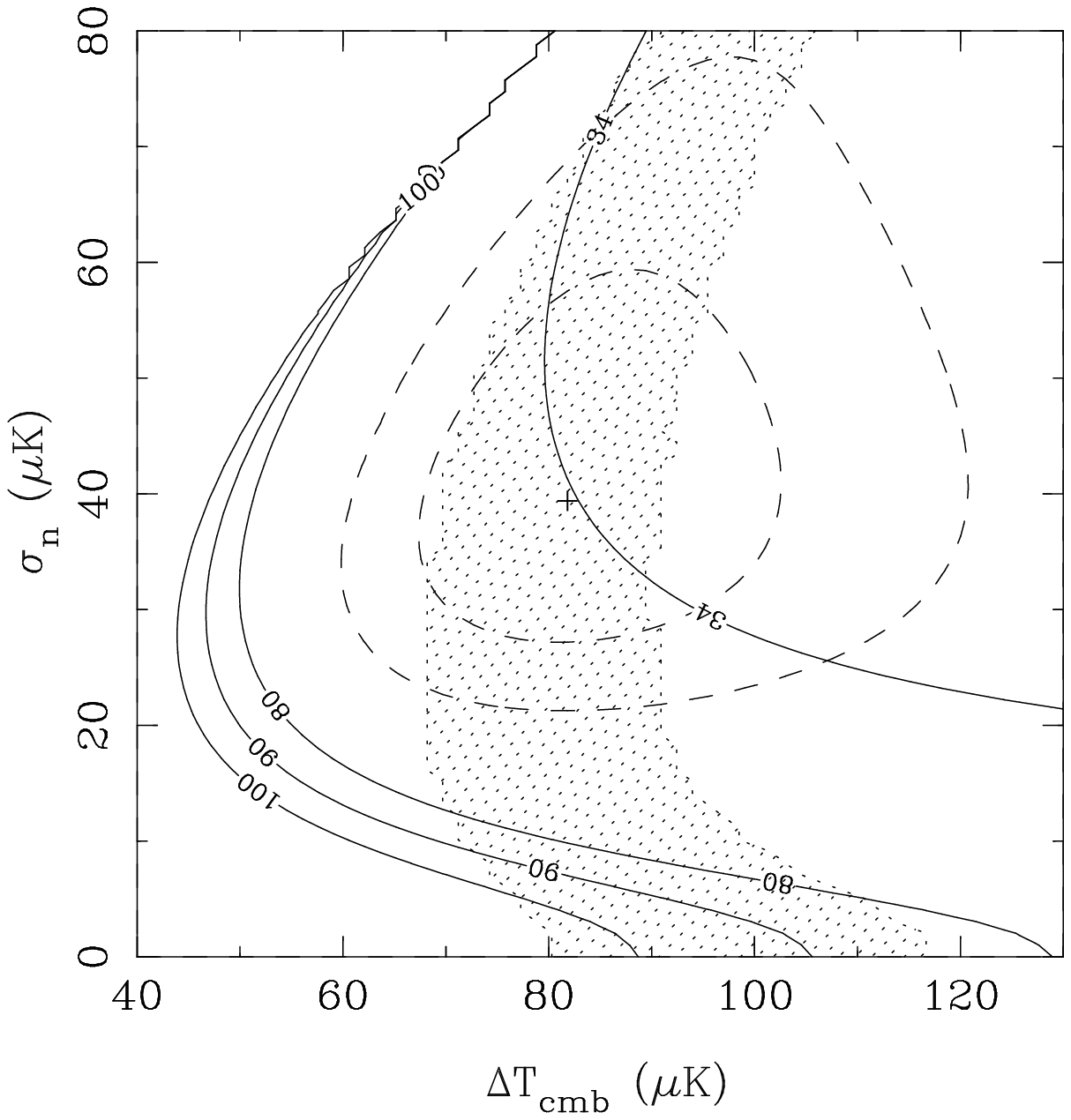}}
\figcaption{\footnotesize Integrated likelihood ${\cal L}(\Delta T_{\rm
cmb}, \sigma_n)$ for the CMBR component extracted at the maximum
likelihood value of $\beta$.  Contours are $\chi^2\equiv{\bf t}^T{\bf
C}^{-1}{\bf t}$ for the 36 Ring5m fields (see
\S\protect\ref{sec:cmbr}).  Dashed contours are the 
$68\%$ and $95\%~\Delta T_{\rm cmb}-\sigma_{\rm n}$ confidence regions.
Stippled region is the locus of 1-D $68\%$ confidence intervals in
$\Delta T_{\rm cmb}$ at each value of $\sigma_{\rm n}$.\label{fig19}}}}
\bigskip

In Figure~\ref{fig19}, we plot ${\cal L}(\Delta T_{\rm cmb},
\sigma_{\rm n})$ for the CMBR component extracted at the maximum
likelihood value of the foreground spectral index ($\beta\simeq
-2.7$), along with contours of constant $\chi^2$, 
where $\chi^2 = {\bf t}^T{\bf C}^{-1}{\bf t}$
(c.f. Eq.~\ref{eq:like}).  The peak of the likelihood function occurs
for $\sigma_{\rm n} = 41^{+20.5}_{-11.8}~\mu$K.  Also shown is the
locus of $68\%$ confidence intervals on $\Delta T_{\rm cmb}$ which
would be obtained if the correlated noise component were held fixed at
the corresponding value on the y-axis.  As can be seen from the
figure, maximum likelihood models which neglect these correlations are
grossly discrepant with the data ($\chi^2 \gg 34$).

\section{Band power}
\label{sec:band}

The mass fluctuation power spectrum is often taken to be
scale-invariant at small wave number, ${\cal P}(k)\propto k$ (i.e., the
Harrison-Zel'dovich spectrum), leading to a CMBR angular power spectrum with
$C^{-1}_l\propto l(l+1)$ on large angular scales
(\cite{Peebles}).  As a result, the {\it band power} (\cite{bondpower})
\begin{equation}
\delta T_l \equiv
\sqrt{l(l+1)C_l/{2\pi}}
\end{equation} 
is expected to be flat at small $l$, and is most often the quantity
predicted by theory.  

With the experimental rms obtained from the diagonal elements of the
covariance matrix in Eq.~\ref{eq:tobs2}, ${\delta T}_{\rm rms}\equiv
\sqrt{{C_{\it ii}}_{\rm cmb}}$, it can be
seen that the band power is related to the {\rm rms} by
\begin{equation}
{\delta T}_{\rm rms} =
\sqrt{\sum_l{{{(l+{1\over{2}})}\over{l(l+1)}}\delta
T^2_l W^{\it ii}_l}},
\end{equation}
so that defining 
\begin{equation}
I(W^{\it ii}_l) = \sum_l{{{(l+{1\over{2}})}\over{l(l+1)}}W^{\it ii}_l},
\end{equation}
the weighted mean of the band power over the window function is given
by
\begin{equation}
\delta T_{l_e} = {\delta T}_{\rm rms}/\sqrt{I(W^{\it ii}_l)}.
\end{equation}
The integral of the window function is $I(W^{\it ii}_l) = 1.96$, yielding
\begin{equation}
\delta T_{l_e} = \dtband, ~l_e = 589^{+167}_{-228},
\label{eq:band}
\end{equation}
where $l_e = I(l W^{\it ii}_l)/I(W^{\it ii}_l)$ and errors on $l_e$
are the points at which the window
function falls by $e^{-0.5}$.  Errors on the band power are the $68\%$
HPD confidence interval, and reflect sample variance, measurement
error and 4.3\% calibration error.

\section{Discussion}

The goal of the RING5M experiment was to determine the CMBR
anisotropy on $7\arcm-22\arcm$ scales, corresponding to $\sim 10$~Mpc
at decoupling.  We detect structure independently at 14.5 and 31.7 GHz well
above the noise limits of the data.  Observation of northern fields
near transit ensures that the local environment is
nearly identical from one field to the next; although the 14.5 GHz data show
a parallactic angle dependence indicative of contamination by RFI,
observations of the fields at both upper and lower culmination confirm that
RFI contributes the same offset to each field and thus does not affect
our estimate of the variance.  Numerous internal consistency checks
(\S\ref{sec:data}-\ref{sec:rfi}) demonstrate that the structure
observed in the RING5M is reproducible
from year to year at both frequencies, while the strong correlation
between frequencies confirms that the structure originates on the
sky (\S\ref{sec:foregrounds}).  Careful characterization of both telescopes
(\S\ref{sec:calibration}) and an extensive
program of observations of calibrator sources
(\S\ref{sec:fluxscale}) reduce our total
calibration error to 4.3\% --- well below the sample variance of the
experiment.

With discrimination of foregrounds provided by our low frequency
channel and accurate removal of point source contamination using the
VLA, we can state with confidence that 76\% of the raw variance, or
88\% of the source-subtracted variance at 31.7 GHz is due to the CMBR
(see Table~\ref{tab7}).  (A breakdown of the 31.7 GHz data set into
CMBR, point source and foreground contributions is given in Table~\ref{tab8}.)

\begin{deluxetable}{lrrrrrrrr}
\footnotesize
\tablewidth{0pt}
\tablenum{8}
\tablecaption{\label{tab8}\sc Summary of Contributions to the 31.7 GHz
Ring Field Means\tablenotemark{a}}
\tablehead{\colhead{Field} & \colhead{$\overline{\Delta T}_{\rm raw}~$($\mu$K)} &\colhead{$\epsilon_{\rm raw}~$($\mu$K)\tablenotemark{b}} & \colhead{$\overline{\Delta T}_{\rm src}~$($\mu$K)} & \colhead{$\epsilon_{\rm src}~$($\mu$K)} & \colhead{$\overline{\Delta T}_{\rm cmb}~$($\mu$K)} & \colhead{$\epsilon_{\rm cmb}~$($\mu$K)} & \colhead{$\overline{\Delta T}_{\rm fore}~$($\mu$K)} & \colhead{$\epsilon_{\rm fore}~$($\mu$K)}}
\startdata
{\sc OV5M0024} & $75.88$ & $17.63$ & $143.66$ & $17.07$ & $72.73$ & $50.34$ & $-140.52$ & $20.38$ \nl
{\sc OV5M0104} & $-118.54$ & $17.73$ & $-146.49$ & $17.28$ & $-52.91$ & $45.35$ & $80.85$ & $19.14$ \nl
{\sc OV5M0144} & $-43.42$ & $17.11$ & $-9.83$ & $0.84$ & $-20.42$ & $21.49$ & $-13.17$ & $5.43$ \nl
{\sc OV5M0224} & $110.99$ & $15.77$ & $20.27$ & $1.15$ & $36.65$ & $20.09$ & $54.08$ & $5.36$ \nl
{\sc OV5M0304} & $-86.21$ & $16.35$ & $-10.60$ & $0.90$ & $-55.68$ & $20.43$ & $-19.93$ & $4.86$ \nl
{\sc OV5M0344} & $-19.82$ & $17.35$ & $1.72$ & $0.06$ & $-21.33$ & $21.64$ & $-0.22$ & $5.19$ \nl
{\sc OV5M0424} & $90.33$ & $16.07$ & $10.75$ & $0.28$ & $114.66$ & $20.45$ & $-35.07$ & $5.71$ \nl
{\sc OV5M0504} & $-13.56$ & $16.32$ & $1.92$ & $0.05$ & $-6.60$ & $20.38$ & $-8.89$ & $4.97$ \nl
{\sc OV5M0544} & $-11.28$ & $16.50$ & $2.62$ & $0.55$ & $-27.46$ & $20.55$ & $13.56$ & $4.82$ \nl
{\sc OV5M0624} & $96.01$ & $15.60$ & $0.05$ & $1.87$ & $100.92$ & $19.72$ & $-4.96$ & $4.96$ \nl
{\sc OV5M0704} & $132.13$ & $15.87$ & $5.30$ & $0.71$ & $174.83$ & $19.84$ & $-48.00$ & $4.77$ \nl
{\sc OV5M0744} & $12.64$ & $15.29$ & $-1.23$ & $0.65$ & $18.66$ & $19.16$ & $-4.79$ & $4.75$ \nl
{\sc OV5M0824} & $-122.33$ & $13.87$ & $1.80$ & $0.29$ & $-73.66$ & $17.33$ & $-50.47$ & $4.21$ \nl
{\sc OV5M0904} & $145.15$ & $14.56$ & $4.62$ & $0.36$ & $99.82$ & $18.16$ & $40.72$ & $4.34$ \nl
{\sc OV5M0944} & $45.37$ & $15.75$ & $-0.87$ & $0.35$ & $94.81$ & $19.68$ & $-48.58$ & $4.67$ \nl
{\sc OV5M1024} & $33.42$ & $14.26$ & $13.68$ & $0.71$ & $5.21$ & $19.73$ & $14.53$ & $6.88$ \nl
{\sc OV5M1104} & $0.56$ & $15.12$ & $8.38$ & $0.69$ & $-67.08$ & $19.19$ & $59.27$ & $4.93$ \nl
{\sc OV5M1144} & $-78.24$ & $15.07$ & $-21.47$ & $1.17$ & $14.13$ & $20.43$ & $-70.89$ & $6.25$ \nl
{\sc OV5M1224} & $109.97$ & $14.63$ & $58.15$ & $2.32$ & $-11.15$ & $23.51$ & $62.97$ & $9.28$ \nl
{\sc OV5M1304} & $-62.25$ & $14.80$ & $-21.74$ & $2.38$ & $-38.47$ & $19.74$ & $-2.03$ & $5.81$ \nl
{\sc OV5M1344} & $-83.76$ & $14.86$ & $18.59$ & $3.58$ & $-79.56$ & $19.29$ & $-22.80$ & $4.84$ \nl
{\sc OV5M1424} & $-37.67$ & $15.06$ & $-28.83$ & $3.61$ & $-34.28$ & $19.93$ & $25.45$ & $5.97$ \nl
{\sc OV5M1504} & $106.89$ & $15.74$ & $48.17$ & $6.14$ & $22.48$ & $22.14$ & $36.24$ & $7.48$ \nl
{\sc OV5M1544} & $-46.12$ & $16.04$ & $-11.72$ & $2.61$ & $-27.78$ & $20.35$ & $-6.63$ & $5.07$ \nl
{\sc OV5M1624} & $-68.42$ & $15.85$ & $-0.91$ & $0.29$ & $-71.48$ & $19.84$ & $3.96$ & $4.93$ \nl
{\sc OV5M1704} & $-48.09$ & $15.69$ & $-9.81$ & $1.45$ & $-33.43$ & $19.70$ & $-4.84$ & $4.85$ \nl
{\sc OV5M1744} & $35.44$ & $16.18$ & $22.33$ & $1.88$ & $-28.20$ & $20.50$ & $41.31$ & $5.26$ \nl
{\sc OV5M1824} & $-126.52$ & $15.40$ & $-11.98$ & $1.70$ & $-102.90$ & $19.47$ & $-11.64$ & $5.03$ \nl
{\sc OV5M1904} & $66.62$ & $16.01$ & $-1.79$ & $0.64$ & $60.59$ & $19.99$ & $7.82$ & $4.80$ \nl
{\sc OV5M1944} & $-39.11$ & $15.28$ & $9.50$ & $0.53$ & $-32.86$ & $19.22$ & $-15.75$ & $4.94$ \nl
{\sc OV5M2024} & $-192.57$ & $16.47$ & $-0.89$ & $0.71$ & $-169.52$ & $20.67$ & $-22.15$ & $5.17$ \nl
{\sc OV5M2104} & $175.60$ & $16.14$ & $-0.97$ & $0.41$ & $157.75$ & $20.30$ & $18.83$ & $5.20$ \nl
{\sc OV5M2144} & $-13.04$ & $16.62$ & $8.55$ & $1.72$ & $-59.85$ & $22.04$ & $38.26$ & $7.05$ \nl
{\sc OV5M2224} & $176.05$ & $16.63$ & $12.56$ & $0.71$ & $149.43$ & $21.40$ & $14.07$ & $5.93$ \nl
{\sc OV5M2304} & $-142.53$ & $17.61$ & $-7.96$ & $0.60$ & $-119.42$ & $21.99$ & $-15.16$ & $5.22$ \nl
{\sc OV5M2344} & $-59.58$ & $18.60$ & $-105.50$ & $12.23$ & $11.36$ & $35.07$ & $34.57$ & $10.66$ \nl
\enddata
\tablenotetext{a}{Equivalent R-J temperature.}
\tablenotetext{b}{Errors are $1\sigma$ rms in the sample mean.}
\end{deluxetable}

Our 14.5 GHz observations of the NCP have also resulted in the detection
of an anomalous component of Galactic emission.  The detection of this
component is a cautionary tale for future CMBR experiments; whether
the emission is due to high-temperature free-free emission correlated
with Galactic dust (Leitch et al. 1997), or to some novel type of
emission from the dust itself (\cite{Ferrara}, \cite{Draine}), our
results suggest that emission associated with Galactic IR cirrus
is potentially a serious contaminant of small-scale anisotropy
experiments even below 100 GHz.

\section{Conclusion}

A likelihood analysis of the RING5M data yields, for the CMBR
component alone, a temperature {\rm rms} of $\delta T_{\rm rms} =
\dtrms$, or equivalently, a band power of $\delta
T_l \equiv \sqrt{l(l+1)C_l/2\pi} = \dtband$
($68\%$ HPD confidence interval) at $l_e = 589^{+167}_{-228}$.

On $2\arcm$ scales, the OVRO NCP 95\% confidence upper limit of
$\Delta T/T < 1.7\times 10^{-5}$ has recently been confirmed by the
SuZIE experiment, which set an upper limit of $\Delta T/T < 2.1\times
10^{-5}$ at nearly the same angular scale (\cite{suzie}).  While these
results cannot strongly differentiate between varieties of closed
universes, as anisotropies are exponentially
damped at arcminute scales for all of these models (\cite{silk}), the
RING5M, NCP and SuZIE results together constitute a significant
constraint on open
universe models; models with $h = 0.3-0.7$ and $\Omega_b = \Omega_0 =
0.2$ over-predict small-scale power by $35\%-50\%$ at $l_{\rm
RING5M}$, and by $10\%-35\%$ over the NCP upper limit at $l_{\rm
NCP}$.  Open models with baryon density close to the lower bound
allowed by big bang nucleosynthesis calculations, $\Omega_bh^2\sim 0.015$
(\cite{BBN1}, \cite{BBN2}), can reproduce the band powers observed at
arcminute scales but severely under-predict the power observed by
degree-scale experiments.  

Taken together, the RING5M and NCP results indicate a decrease
in the angular power spectrum between $l\sim 600$ and $l\sim 2000$.  
If the collection of data near $l$ of a few hundred can be taken as
evidence for a rise in the power spectrum toward small scales, then the
RING5M band power is consistent with a peak in the power spectrum near
$l\sim 200$.  A $\chi^2$ fit to the data in
Figure~\ref{fig19} for a range of model power spectra shows that
the data are consistent with $\Omega_0 = 1$ in a $\Lambda$-model with
$\Omega_bh^2 = 0.015$ and $\Omega_\Lambda = 0.7$ or a standard CDM
scenario with $\Omega_bh^2 = 0.0045$, shown in
Figure~\ref{fig19}.  Although increasing $\Omega_b$
dramatically affects the amplitude of the first acoustic peak, the
competing effects of damping at small scales and enhancement of
compressional peaks with increasing $\Omega_b$ means that the RING5M
result cannot strongly constrain $\Omega_b$ in a flat universe.

Recent observations with the Cambridge Anisotropy
Telescope (CAT) have resulted in a detection of anisotropy on
angular scales directly comparable to those probed by the RING5M
experiment (\cite{CAT2}).  The broadband power reported
in this paper is in good agreement with the CAT detection of 
$\delta T_{l_{e}}/T = 1.8^{+0.7}_{-0.5}\times 10^{-5}$ at $l = 590$.
Given the result reported here, and the NCP and SuZIE results at
higher $l$, there can be
little doubt that the CMBR spectrum drops significantly between
$l\sim 600$ and $l\sim 2000$, as expected in flat cosmological models.

\acknowledgments

We are indebted to Harry Hardebeck, Mark
Hodges and Russ Keeney for their consistently exceptional work on the 5.5-meter
and 40-meter telescopes and receivers.  We thank Marion Pospiezalski
at NRAO and Javier Bautista at JPL for providing the HEMT amplifiers
upon which this experiment is based.  We thank John Carlstrom for the
loan of a HEMT amplifier, and Bharat Ratra and Ken Ganga for pointing
out an error in the manuscript.  We are grateful to U. Seljak and
M. Zaldarriaga for placing CMBFAST in
the public domain.  The work at the California Institute of Technology
was supported by NSF grants AST-9119847 and AST-9419279.  The VLA
is an instrument of the National Radio Astronomy Observatory, a
facility of the National Science Foundation operated under cooperative
agreement by Associated Universities, Inc. 

\onecolumn

\centerline{\vbox{\epsfxsize=15.0cm\center{\hspace{0.15cm}\epsfbox{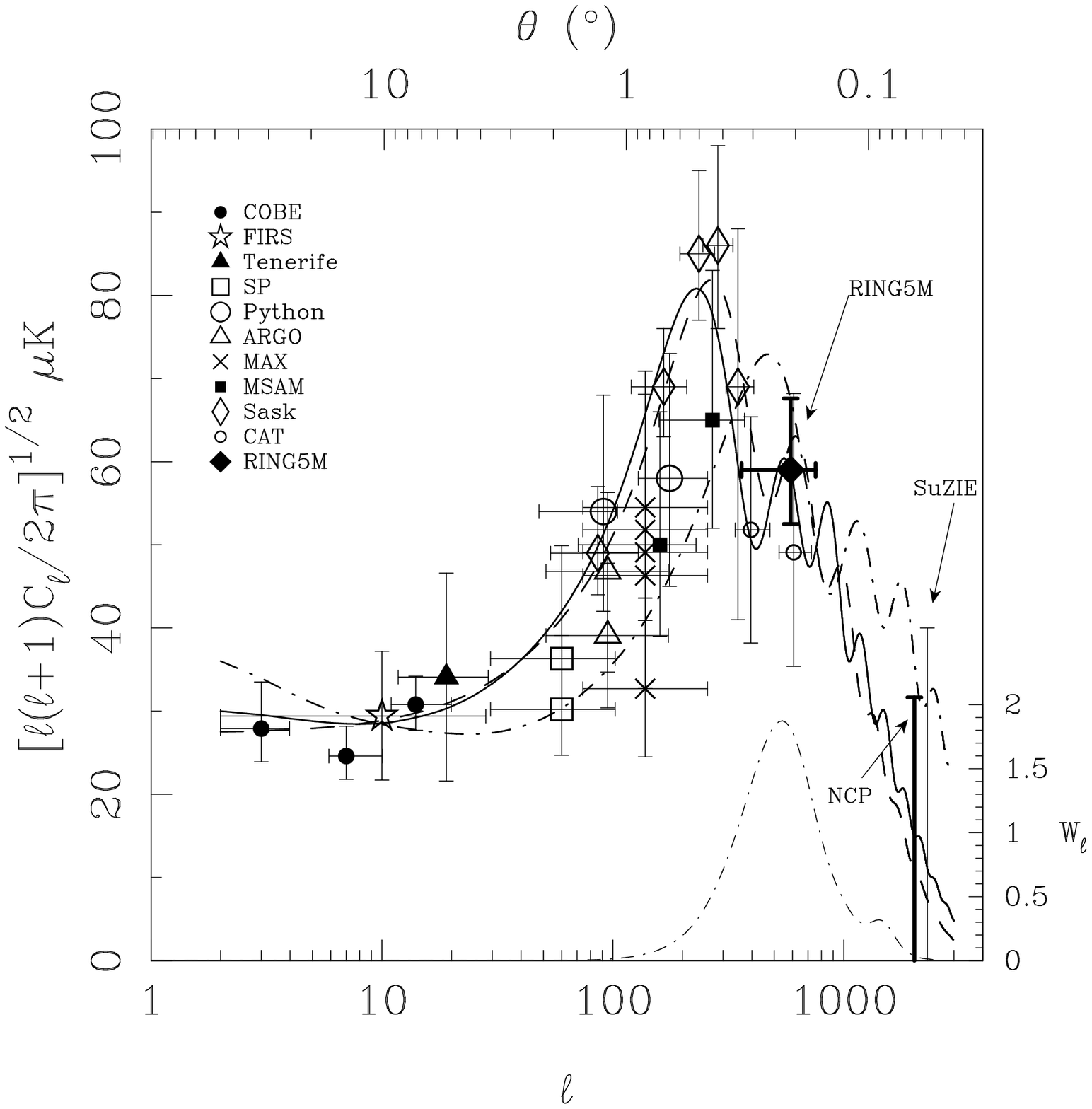}}
\figcaption{\footnotesize CMBR anisotropy measurements, shown with a
$\Lambda$- model (solid line) with $\Omega_bh^2 = 0.015,
\Omega_\Lambda = 0.7$, and $\Omega_0 = 1$, a standard CDM model
(dashed line) with $\Omega_bh^2 = 0.0045$, and an open model
(dot-dashed line) with $\Omega_bh^2 = 0.015$ and $\Omega_0 = 0.3$.  At
bottom is the RING5M zero-lag window function (see \S\ref{sec:wfn}).
Indicated in bold are the OVRO RING5M detection of anisotropy at $l_e
= 589$ (this paper) and the 95\% confidence upper limit from the OVRO
NCP experiment (\cite{NCP}).  Other data points are: COBE
(\cite{Hinshaw}), FIRS (\cite{Ganga}), Tenerife (\cite{ten}), SP
(\cite{SP}), Python (\cite{python2}), ARGO (\cite{argo}), MAX
(\cite{MAX2}), Saskatoon (\cite{sask}), CAT (\cite{CAT2}) and SuZIE
(\cite{suzie}).  Models were computed using CMBFAST (\cite{cmbfast1},
\cite{cmbfast2}, \cite{cmbfast3}).}
\label{fig20}}}

\end{document}